\documentclass[prd,preprint,tightenlines,floatfix,showpacs,preprintnumbers,nofootinbib,eqsecnum,superscriptaddress]{revtex4}

\usepackage[dvips,final]{graphicx}
 \usepackage{amssymb}
  \usepackage{amsmath}
   \usepackage{amsfonts}
    \usepackage{epsfig}
     \usepackage{bm}% bold math
      \usepackage{multirow}

     \newcommand{\bp}{\mbox{\boldmath $p$}}
     \newcommand{\bq}{\mbox{\boldmath $q$}}
     \newcommand{\bk}{\mbox{\boldmath $k$}}
\begin{document}

\title{
Measurement of exclusive production of scalar
\mbox{\boldmath $\chi_{c0}$} meson\\
in proton-(anti)proton collisions via
\mbox{\boldmath $\chi_{c0} \to \pi^{+}\pi^{-}$} decay}

\author{P. Lebiedowicz}
\email{piotr.lebiedowicz@ifj.edu.pl}
\affiliation{Institute of Nuclear Physics PAN, PL-31-342 Cracow, Poland}

\author{R. Pasechnik}
\email{roman.pasechnik@fysast.uu.se}
\affiliation{Department of Physics and Astronomy, Uppsala University, Box 516, SE-751 20 Uppsala, Sweden}

\author{A. Szczurek}
\email{antoni.szczurek@ifj.edu.pl}
\affiliation{University of Rzesz\'ow, PL-35-959 Rzesz\'ow, Poland}
\affiliation{Institute of Nuclear Physics PAN, PL-31-342 Cracow, Poland}

\begin{abstract}
We consider a measurement of exclusive production of scalar
$\chi_{c}(0^{++})$ meson
in the proton-proton collisions at LHC and RHIC
and in the proton-antiproton collisions at the Tevatron
via $\chi_{c0} \to \pi^{+}\pi^{-}$ decay.
The corresponding amplitude for exclusive
double-diffractive $\chi_{c0}$ meson production
was obtained within the $k_{t}$-factorization approach
including virtualities of active gluons
and the corresponding cross section is calculated with
unintegrated gluon distribution functions (UGDFs)
known from the literature.
The four-body $p p \to p p \pi^+ \pi^-$ reaction constitutes
an irreducible background to the exclusive
$\chi_{c0}$ meson production.
We calculate several differential distributions
for $pp(\bar{p}) \to pp(\bar{p})\chi_{c0}$
process including absorptive corrections.
The influence of kinematical cuts on the signal-to-background ratio
is investigated. Corresponding experimental consequences are
discussed.
\end{abstract}

%Keywords:

\maketitle

%--------------------------------------------------
\section{Introduction}
%--------------------------------------------------
The mechanism of exclusive production of mesons at high energies
became recently a very active field of research (see e.g.
Ref.~\cite{ACF10} and references therein). Central exclusive
production processes represent a very promising and novel way to
study QCD in hadron-hadron collisions. Recently, there is a growing
interest in understanding exclusive three-body reactions $p p \to p
p M$ at high energies, where the meson (resonance) $M$ is produced
in the central rapidity region. In particular, these reactions
provide a valuable tool to investigate in detail the properties of
resonance states. Many of the resonances decay into $\pi\pi$ and/or
$KK$ channels. The representative examples are: $M=\sigma, \rho^{0},
f_{0}(980), \phi, f_{2}(1275), f_{0}(1500)$ and, as we shall
emphasis here, $\chi_{c0}$. Various decay channels can be studied.
It is clear that these resonances are seen (or will be seen) "on"
the background of a $\pi\pi$ or $KK$ continuum \footnote{In general,
the resonance and continuum contributions may interfere. This may
produce even a dip. A good example is the $f_{0}(980)$ production
(see e.g. Ref.~\cite{LSK09,Alde97}).}. The two-pion background to
exclusive production of $f_{0}(1500)$ meson was already discussed in
Ref.~\cite{SL09}. The recent works concentrated on the production of
$\chi_c$ mesons (see e.g.
Refs.~\cite{PST_chic0,PST_chic1,PST_chic2,LKRS10,LKRS11} and
references therein) where the QCD mechanism is similar to the
exclusive production of the Higgs boson. Furthermore, the
$\chi_{c(0,2)}$ states are expected to annihilate via two-gluon
processes into light mesons and may, therefore, allow the study of
glueball production dynamics \cite{glueballs}.

Recently, the CDF Collaboration has measured the cross section for
exclusive production of $\chi_c$ mesons in proton-antiproton
collisions at the Tevatron \cite{CDF_chic}, by selecting events with
large rapidity gaps separating the centrally produced state from the
dissociation products of the incoming protons. In this experiment
$\chi_c$ mesons are identified via decay to the $J/\psi + \gamma$
with $J/\psi \to \mu^{+}\mu^{-}$ channel. The experimental invariant
mass resolution was not sufficient to distinguish between scalar,
axial and tensor $\chi_c$. While the branching fractions to this
channel for axial and tensor mesons are large \cite{PDG}
($\mathcal{B} = (34.4 \pm 1.5)\%$ and $\mathcal{B} = (19.5 \pm 0.8)\%$, respectively)
the branching fraction for the scalar meson is very small 
$\mathcal{B} = (1.16 \pm 0.08 )\%$ \cite{PDG}. 
On the other hand, the cross section
for exclusive $\chi_{c0}$ production obtained within the
$k_{t}$-factorization is much bigger than that for $\chi_{c1}$ and
$\chi_{c2}$. As a consequence, all $\chi_{c}$ mesons give similar
contributions \cite{PST_chic2} to the $J/\psi + \gamma$ decay
channel. Clearly, the measurement via decay to the $J/\psi + \gamma$
channel cannot provide cross section for different $\chi_c$.

Could other decay channels be used?
The $\chi_{c0}$ meson decays
into several two-body channels (e.g. $\pi \pi$, $K^{+} K^{-}$, $p \bar{p}$)
or four-body hadronic modes (e.g. $\pi^{+} \pi^{-} \pi^{+} \pi^{-}$,
%$\pi^{+} \pi^{-} \pi^{0} \pi^{0}$,
$\pi^{+} \pi^{-} K^{+} K^{-}$). The branching ratios are shown in
Table~\ref{tab:hadronic_modes}. In this paper we analyze a
possibility to measure $\chi_{c0}$ via its decay to $\pi^+ \pi^-$
channel. The advantage of this channel is that the $\pi^+ \pi^-$
continuum has been studied recently \cite{LS10} and is relatively
well known. In addition the axial $\chi_{c1}$ does not decay to the
$\pi \pi$ channel and the branching ratio for the $\chi_{c2}$ decay
into two pions is smaller.
%$\mathcal{B}(\chi_{c2} \to \pi\pi) = (2.39 \pm 0.14) \times 10^{-3}$ \cite{PDG}.
A much smaller cross section for $\chi_{c2}$ production means, in
practice, that only $\chi_{c0}$ will contribute to the signal.

%=======================================================================================
\begin{table}[h]
\caption{Branching fractions for the $\chi_{cJ}$ two- and four-body
hadronic decays, taken from Ref.~\cite{PDG}.}
\label{tab:hadronic_modes}
\begin{center}
\begin{tabular}{|c|c|c|c|}
\hline\hline
Channel & $\mathcal{B}(\chi_{c0})$ & $\mathcal{B}(\chi_{c1})$ & $\mathcal{B}(\chi_{c2})$  \\
\hline
$\pi^{+} \pi^{-}$   & $(0.56 \pm 0.03)\times 10^{-2}$   & $-$ & $(0.16 \pm 0.01)\times 10^{-2}$\\
$K^{+} K^{-}$       & $(0.610 \pm 0.035)\times 10^{-2}$ & $-$ & $(0.109 \pm 0.008)\times 10^{-2}$\\
$p\bar{p}$          & $(2.28 \pm 0.13)\times 10^{-4}$   & $(0.73 \pm 0.04)\times 10^{-4}$&$(0.72 \pm 0.04)\times 10^{-4}$\\
$\pi^{+} \pi^{-} \pi^{+} \pi^{-}$ & $(2.27 \pm 0.19)\times 10^{-2}$ & $(0.76 \pm 0.26)\times 10^{-2}$ & $(1.11 \pm 0.11)\times 10^{-2}$\\
%$\pi^{+} \pi^{-} \pi^{0} \pi^{0}$ & $(3.4 \pm 0.4)\times 10^{-2}$  & $(1.26 \pm 0.17)\times 10^{-2}$ & $(2.00 \pm 0.26)\times 10^{-2}$\\
$\pi^{+} \pi^{-}   K^{+}   K^{-}$ & $(1.80 \pm 0.15)\times 10^{-2}$ & $(0.45 \pm 0.10)\times 10^{-2}$& $(0.92 \pm 0.11)\times 10^{-2}$\\
\hline\hline
\end{tabular}
\end{center}
\end{table}
%=====================================================================

In the present paper, we wish to calculate differential
distributions for the exclusive production $\chi_{c0}$ meson with a
few UGDFs taken from the literature relevant for small gluon
virtualities (transverse momenta). We shall use matrix element for
the off-shell gluons as obtained in Ref.~\cite{PST_chic0}. The
expected non-resonant background can be modeled using a
"non-perturbative" framework, mediated by pomeron-pomeron fusion
with an intermediate off-shell pion exchanged between the
final-state particle pairs. Thus, we consider $p p(\bar{p}) \to p
p(\bar{p}) \pi^{+}\pi^{-}$ reaction as a genuine four-body process
with exact kinematics which can be easily used when kinematical
cuts have to be improved.

Exclusive charmonium decays have been a subject of interest at the
$e^{+}e^{-}$ colliders as they are an excellent laboratory for
studying quark-gluon dynamics at relatively low energies. 
Thus, a measurement of many exclusive hadronic $\chi_c$ decays if possible
is very valuable. Although these $\chi_{c}$ states are not directly
produced in $e^{+}e^{-}$ collisions, they are copiously produced in
the radiative decays $\psi(2S) \to \gamma \chi_{c}$, each of which
has a branching ratio of around 9\% \cite{PDG}.
%On the other hand these $\chi_{c}$ states can be produced
%in the radiative decays $\psi' \to \gamma \chi_{c}$,
%providing a clean enviroment to study their decays.
The CLEO Collaboration has studied exclusive $\chi_{c(0,1,2)}$
decays to four-hadron final states involving two charged and two
neutral mesons \cite{CLEO}: $\pi^{+}\pi^{-}\pi^{0}\pi^{0}$,
$K^{+}K^{-}\pi^{0}\pi^{0}$, $p \bar{p}\pi^{0}\pi^{0}$,
$K^{+}K^{-}\eta\pi^{0}$ and $K^{\pm}\pi^{\mp}K^{0}\pi^{0}$. 
The BESIII Collaboration has studied two-body $\chi_{c(0,2)}$ decays
into $\pi^0\pi^0$ and $\eta\eta$ \cite{BES_pi0pi0_etaeta}
\footnote{The $\chi_{c1}$ decay into these final states are not
considered as they are forbidden by the spin-parity conservation.}
and four-body $\chi_{c(0,1,2)}$ decays into
$\pi^{0}\pi^{0}\pi^{0}\pi^{0}$ \cite{BES_4pi0} final states,
where $\chi_{cJ}$ signals appear in radiative photon energy spectrum.
Recently (Ref.~\cite{BES_ppKK}) a study of $\chi_{cJ}$ hadronic decays 
and measurements of $\chi_{cJ} \to \Lambda(1520) \bar{\Lambda}(1520)$ decaying
into $p \bar{p} K^{+} K^{-}$ has been presented.
In the present paper we start
the discussion of the possibility to measure the different decay
channels in proton-(anti)proton collisions in order to determine the cross
section for exclusive production of the P-wave charmonia. Here,
continuum backgrounds can be larger than in the $e^+ e^-$ collisions and
this requires a separate discussion of the feasibility. We will
discuss this issue in the present paper.

The paper is organized as follows.
In section \ref{section:II} we give general expressions
for the amplitudes of the exclusive $\chi_{c0}$ meson production
and $\pi^{+}\pi^{-}$ pairs production
including discussion of absorptive corrections.
Section \ref{section:III} contains the presentation
of the main results and a discussion of the uncertainties related to
the approximations made.
Finally, some concluding remarks and outlook are given
in Section \ref{section:IV}.

%-------------------------------
\section{Signal and background amplitudes}
\label{section:II}
%-------------------------------
%--------------------------------------
\subsection{Diffractive QCD amplitude for exclusive \mbox{\boldmath $\chi_{c0}$} production}
\label{subsection:signal}
%--------------------------------------
%--------------------------------------------------------------------
\begin{figure}[!h]    % Figure 1
 \centerline{\includegraphics[width=0.4\textwidth]{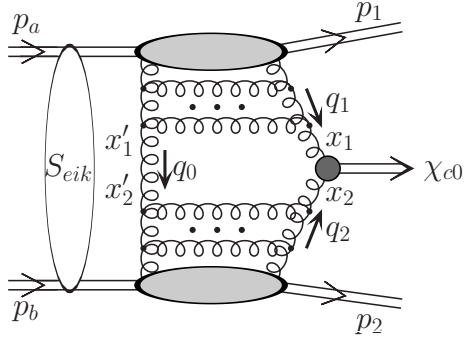}}
   \caption{\label{fig:fig1}
   \small  The QCD mechanism of exclusive diffractive production of $\chi_{c0}$ meson
   including the absorptive correction.}
\end{figure}
%--------------------------------------------------------------------
The QCD mechanism for the diffractive production of heavy central
system has been proposed by Khoze, Martin and Ryskin (KMR) and
developed in collaboration with Kaidalov and Stirling for Higgs
production (see e.g. Refs.~\cite{KMR,KKMR,KKMR-spin,KMRS}).
In the framework of this
approach the amplitude of the exclusive $pp\to pp \chi_{c0}$ process
is described by the diagram shown in Fig.~\ref{fig:fig1}, where the
hard subprocess $g^{*}g^{*} \to \chi_{c0}$ is initiated by the
fusion of two off-shell gluons and the soft part represented in
terms of the off-diagonal unintegrated gluon distributions (UGDFs).
The formalism used to calculate the exclusive $\chi_{c0}$ meson
production is explained in detail elsewhere \cite{PST_chic0} and so
we will only review relevant aspects here.

The "full" amplitude for the exclusive process $pp \to pp\chi_{c0}$
can be written as
\begin{eqnarray}
{\cal {M}}_{pp \to pp\chi_{c}}^{full}(s,y,-\bp_{1,t},-\bp_{2,t}) &=&  {\cal {M}}_{pp \to pp\chi_{c}}^{bare}(s,y,-\bp_{1,t},-\bp_{2,t})\nonumber\\
&+&{\cal {M}}_{pp \to pp\chi_{c}}^{rescatt}(s,y,-\bp_{1,t},-\bp_{2,t})\, .
\label{amp_full}
\end{eqnarray}
We can write the "bare" amplitude \cite{PST_chic0} as
\begin{eqnarray}
{\cal M}_{pp \to pp\chi_{c}}^{bare}(s,y,-\bp_{1,t},-\bp_{2,t})=\nonumber\\
\frac{s}{2} \pi^2 \frac12 \frac{1}{N_c^2-1}
 \Im \int d^2 q_{0,t} V(\bq_{1,t},\bq_{2,t})
\frac{f^{off}_{g1}(x_1,x',q_{0,t}^2,q_{1,t}^2,t_1)
      f^{off}_{g2}(x_2,x',q_{0,t}^2,q_{2,t}^2,t_2)}
     {q_{0,t}^{2}\,q_{1,t}^{2}\, q_{2,t}^{2}} \,,
\label{amp_bare}
\end{eqnarray}
where the objects $f_{g1/g2}^{off}$ are skewed
(or off-diagonal) unintegrated gluon distributions of both nucleons.
$t_{1,2}$ are the momentum transfers along each nucleon line
\footnote{In the following for brevity we shall use notation $t_{1,2}$
which means $t_1$ or $t_2$.},
$q_{1,t}$, $q_{2,t}$, $q_{0,t}$, $x_{1,2}$,
$x'_{1} \sim x'_{2} \ll x_{1,2}$
are the transverse momenta and the longitudinal momentum fractions
for active and screening gluons, respectively.
UGDFs are nondiagonal both in the $x$ and $q_{t}^{2}$ space.
The usual off-diagonal gluon distributions are nondiagonal only in $x$.
In the limit $x_{1,2} \to x'_{1,2}$, $q_{0,t}^{2} \to q_{1/2,t}^{2}$
and $t_{1,2} \to 0$ they become the usual UGDFs.

The vertex factor $V(\bq_{1,t},\bq_{2,t})$ describes the coupling
of two virtual gluons to $\chi_{c0}$ meson
is obtained in heavy quark approximation and can be written as
\begin{eqnarray}
V(\bq_{1,t},\bq_{2,t}) = K_{NLO} \frac{8i g_{s}^{2}}{M}
\frac{{\cal R}'(0)}{\sqrt{\pi M N_{c}}}
\frac{3 M^{2} \bq_{1,t} \bq_{2,t} -
      2 \bq_{1,t}^{2}\bq_{2,t}^{2}-
      (\bq_{1,t}\bq_{2,t})(\bq_{1,t}^{2} + \bq_{2,t}^{2})}
{(M^{2} + \bq_{1,t}^2 + \bq_{2,t}^2)^{2}}\,,
\label{vertex}
\end{eqnarray}
where $M$ is the $\chi_{c0}$ mass, $g_{s}^{2}=4 \pi
\alpha_{s}(M^{2})$ and the strong coupling constant is calculated in
the leading order and extended to the nonperturbative region
according to Shirkov-Solovtsov analytical model \cite{SS97}. The
value of the $P$-wave radial wave function at the origin is taken to be
\cite{EQ95} ${\cal R}_{\chi_{cJ}}'(0) = \sqrt{0.075}$ GeV$^{5/2}$
and the radiative corrections factor in the vertex $K_{NLO}$ is
well-known \cite{KNLO}, $K_{NLO} \simeq 1.68$.

The rescattering correction shown in Fig.~\ref{fig:fig1} 
by the extra blob can be written in the form
\begin{eqnarray}
{\cal M}_{pp \to pp\chi_{c}}^{rescatt}(s,y,-\bp_{1,t},-\bp_{2,t})&=&
\int \frac{d^2 \bk_{t}}{2(2\pi)^{2}} \frac{A_{pp}(s,k_{t}^{2})}{s}
{\cal M}_{pp \to pp\chi_{c}}^{bare}(s,y,\bk_{1,t},\bk_{2,t})\, ,
\label{amp_rescatt}
\end{eqnarray}
where $\bk_{1,t} = -\bp_{1,t} - \bk_{t}$ and $\bk_{2,t} = -\bp_{2,t}
- \bk_{t}$ with momentum transfer $\bk_{t}$. The amplitude for
elastic proton-proton scattering at an appropriate energy is
conveniently parameterized as
\begin{eqnarray}
A_{pp}(s,k_{t}^{2})&=&
A_{0}(s) \, \exp(-B k_{t}^{2} /2)\, .
\label{pp_scatt}
\end{eqnarray}
From the optical theorem we have Im$A_{0}(s,t = 0) = s \sigma_{tot}(s)$
(the real part is small in the high energy limit).
$B$ is the effective slope of the elastic differential cross section
\begin{equation}
B(s) = B_{i} + 2 \alpha'_{i}
 \ln \left( \frac{s}{s_0} \right) \, ,
\label{slope_NN}
\end{equation}
and is adjusted to the existing experimental data for the elastic $N
N$ scattering. The Donnachie-Landshoff parametrization \cite{DL92}
of the total $pp$ or $p\bar{p}$ cross sections can be used to
calculate the rescattering amplitude. We take $s_{0}$ = 1 GeV$^{2}$,
$\alpha'_{I\!\!P}$ = 0.25 GeV$^{-2}$ 
and $B_{i}$: $B^{NN}_{I\!\!P}$ = 9 GeV$^{-2}$.

The KMR UGDFs, unintegrated over $q_{t}$, are calculated from the
conventional (integrated) distributions $g(x,q_{t}^2)$ and the
so-called Sudakov form factor $\sqrt{T_{g}(q_{t}^{2}, \mu^{2})}$ as
follows:
\begin{eqnarray}
f_{g}^{KMR}(x,x',Q_{t}^2,\mu^{2};t)=
R_{g} \frac{\partial}{\partial\ln q_{t}^2}
\left[
x g(x,q_{t}^2) \sqrt{T_{g}(q_{t}^{2}, \mu^{2})}
\right]
_{q_{t}^{2}=Q_{t}^{2}}
F(t)\, .
% \nonumber\\
%f_{g2}^{KMR}(x_{2},Q_{2,t}^2,\mu^{2};t_{2})&=&
%R_{g} \frac{\partial}{\partial\ln q_{t}^2}
%\left[
%x_{2} g(x_{2},q_{t}^2) \sqrt{T_{g}(q_{t}^{2}, \mu^{2})}
%\right]
%_{q_{t}^{2}=Q_{2,t}^{2}}
%F(t_{2})\, .
\label{fg_kmr}
\end{eqnarray}
The Sudakov factor suppresses real emissions from the active gluon
during the evolution, so that the rapidity gaps survive. The factor
$R_g$ approximately accounts for the single $\mathrm{ln}Q_{t}^{2}$
skewed effect \cite{SGBMR99}. In the calculations presented here we
take $R_g = 1.3$, and value of the hard scale is $\mu^2 = M^{2}/4$.
The choice of the scale is somewhat arbitrary, and consequences of
this choice were discussed in Ref.~\cite{PST_chic0}.

In the original KMR approach the following prescription
for the effective transverse momentum is taken:
\begin{eqnarray}
Q_{1,t}^{2}= \mathrm{min}\left(q_{0,t}^{2}, q_{1,t}^{2}\right)\, , \qquad
Q_{2,t}^{2}= \mathrm{min}\left(q_{0,t}^{2}, q_{2,t}^{2}\right)\, .
\end{eqnarray}
Other prescriptions are also possible \cite{PST_chic0,PST_chic1}. In
the KMR approach only one effective gluon transverse momentum is
taken explicitly in their skewed UGDFs compares to two independent
transverse momenta in our case (see Eq.~(\ref{amp_bare})). Please
note also that the skewed KMR UGDFs does not explicitly depend on
$x'$ (assuming $x' \ll x \ll 1$).

In Ref.~\cite{MR01} a procedure was presented which allows to calculate
the generalized (or skewed) parton distributions of the proton,
$\overline{H}(x, \xi; q_t^2, \mu^2)$, unintegrated over the partonic
transverse momenta, from the conventional parton
distributions, $q (x, \mu^2)$ and $g (x, \mu^2)$, for small values
of the skewedness parameter $\xi^2 \ll 1$ and any $x$. The momentum
fractions carried by the emitted and absorbed partons are defined as
$x_{1,2} = x \pm \frac{1}{2} \xi$ with support $-1 \leq x \leq 1$.
The result is a simple approximate phenomenological form for the distribution:
\begin{eqnarray}
\overline{H}_g \left( \frac{\xi}{2}, \xi; q_t^2, \mu^2 \right) & = &
\sqrt{T_{g}(q_{t}^{2}, \mu^{2})} \left [R_g \:
\frac{\partial xg (x, q_t^2)}{\partial \ln q_t^2} \: + \: xg (x, q_t^2) \:
\frac{N_{c} \alpha_s}{2 \pi} \left (\ln \frac{\mu +
\frac{1}{2} q_t}{q_t} \: + \: 1.2 \frac{\mu^2}{\mu^2 + q_t^2} \right ) \right . \nonumber \\
& & + \; 5 \left . \frac{\alpha_s}{2 \pi} \left (x u_{\rm val}
(x, q_t^2) + xd_{\rm val} (x, q_t^2) \right )
\right ]_{x = \xi} \,.
\label{phenom_form}
\end{eqnarray}

In evaluating $f_{g}$'s we have used the GRV NLO \cite{GRV} and GJR
NLO \cite{GJR} collinear gluon distributions, which allow to use
rather low values of gluon transverse momenta $Q_{t}^2 = q_{0,t}^2$,
$q_{1,t}^2$, $q_{2,t}^2 \geq$ 0.5 GeV$^2$. The collinear
distributions such as CTEQ and MRST are defined for higher
factorization scales ($Q_{t}^2 > 1$ GeV$^2$), and therefore are less
useful in applications discussed here.

The $t$-dependence of the unintegrated gluon distribution
$f_{g}$'s is not well known and is
isolated in the effective form factors of the QCD Pomeron-proton vertex, 
which are parameterized in the
forward scattering limit in the exponential form as
\begin{eqnarray}
F(t)= \exp(b\, t/2)
\label{ff_exp}
\end{eqnarray}
with the $t$-slope parameter $b = 4$ GeV$^{-2}$. 
Then the integral in
Eq.~(\ref{amp_rescatt}) can be evaluated as \cite{KMR_eik}
\begin{eqnarray}
{\cal M}_{pp \to pp\chi_{c}}^{rescatt}(s,y,-\bp_{1,t},-\bp_{2,t})=
\frac{i A_{0}}{4 \pi s (B + 2b)}
\exp \left(
\frac{b^{2} |\bp_{1,t} - \bp_{2,t}|^{2}}{2(B+2b)}
\right)
{\cal M}_{pp \to pp\chi_{c}}^{bare}(s,y,-\bp_{1,t},-\bp_{2,t})\, .
\nonumber \\
\label{amp_rescatt_calc}
\end{eqnarray}

The cross section for the three-body reaction is calculated as
\begin{eqnarray}
\sigma =\int \frac{1}{2s} \overline{ |{\cal M}|^2} (2 \pi)^4
\delta^4 (p_a + p_b - p_1 - p_2 - p_3)
\frac{d^3 p_1}{(2 \pi)^3 2 E_1}
\frac{d^3 p_2}{(2 \pi)^3 2 E_2}
\frac{d^3 p_3}{(2 \pi)^3 2 E_3}
\;
\label{dsigma_for_2to4}
\end{eqnarray}
by choosing convenient kinematical variables.

%--------------------------------------
\subsection{Diffractive amplitude for
\mbox{\boldmath $\pi^{+} \pi^{-}$} continuum}
\label{subsection:baskground}
%--------------------------------------
%--------------------------------------------------------
\begin{figure}[!h]
\includegraphics[width=0.25\textwidth]{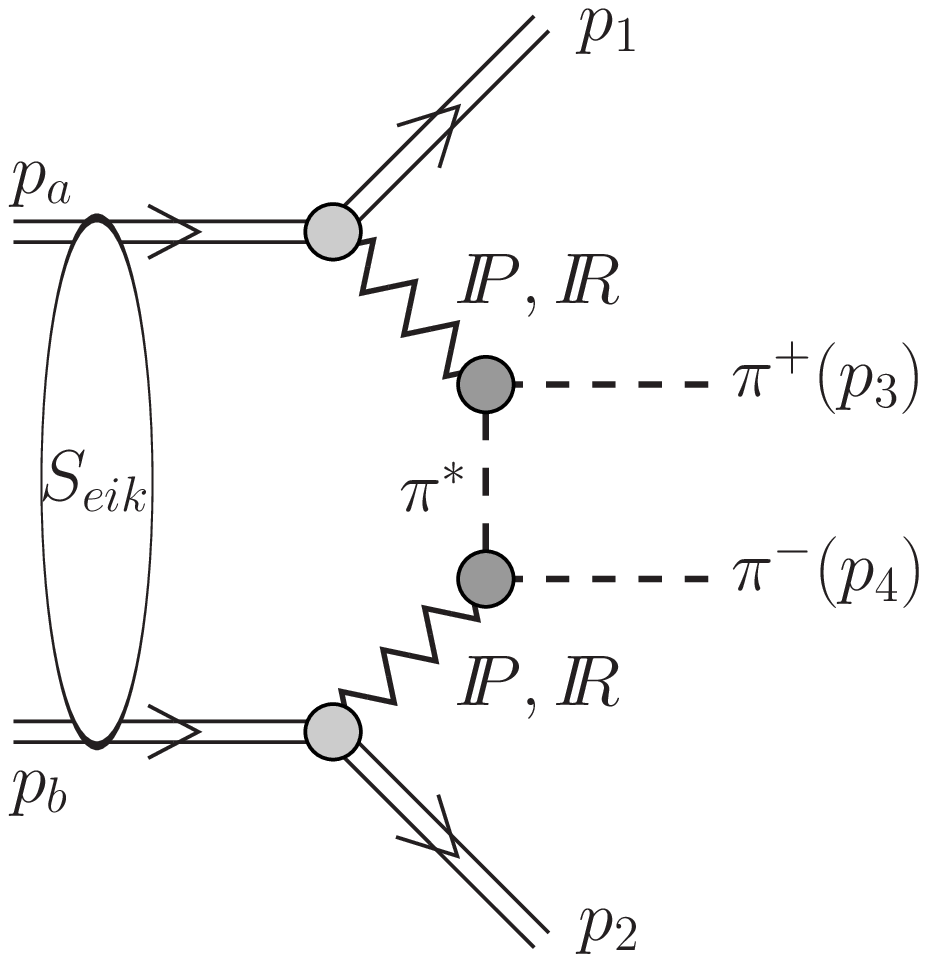}
\includegraphics[width=0.25\textwidth]{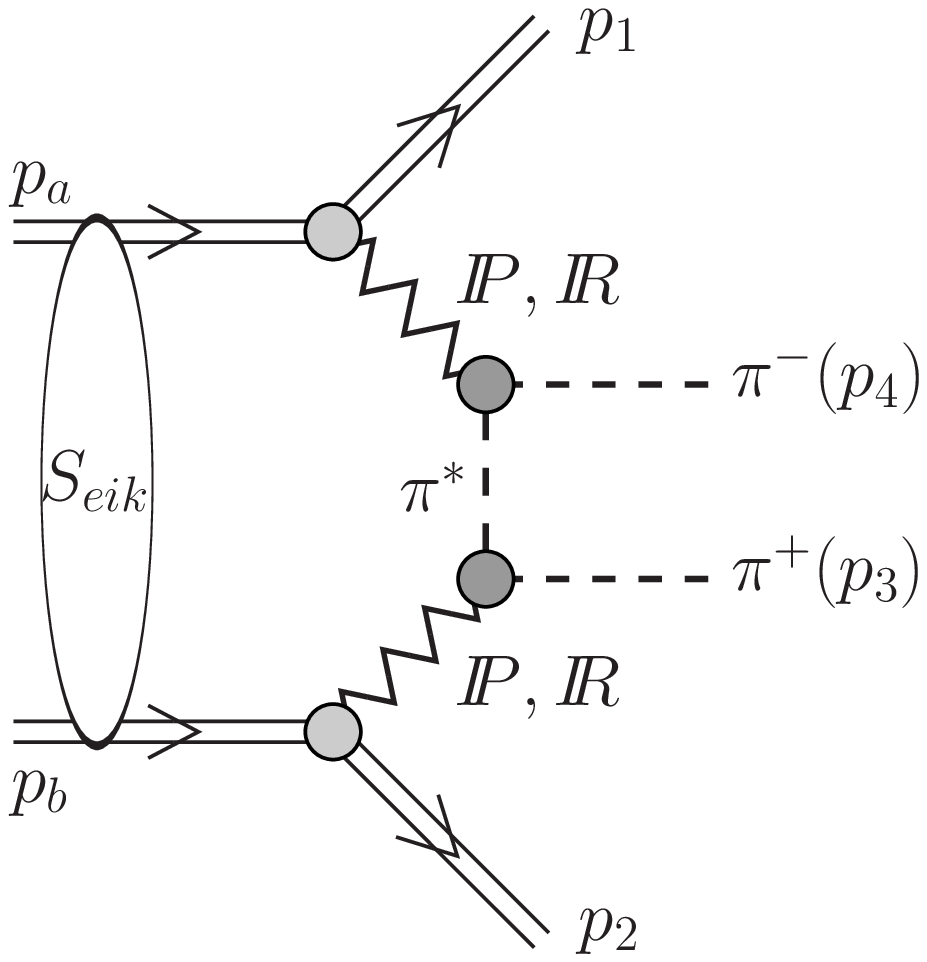}
   \caption{\label{fig:central_double_diffraction_diagrams}
   \small
The double-diffractive mechanism of exclusive production of
$\pi^{+}\pi^{-}$ pairs including the absorptive corrections. }
\end{figure}
%--------------------------------------------------------

The dominant mechanism of the exclusive production of
$\pi^{+}\pi^{-}$ pairs at high energies is sketched in
Fig.~\ref{fig:central_double_diffraction_diagrams}. 
In calculations of the amplitude related to double diffractive mechanism 
for the $pp \to p p \pi^+ \pi^-$ reaction we follow the general rules of
Pumplin and Henyey \cite{PH76} used recently in Ref.~\cite{LS10}
\footnote{For early rough estimates see Ref.~\cite{AKLR75}.
There have also been a variety of experimental results,
in particular from the CERN ISR (for recent reviews see \cite{ACF10}).}.

The "full" amplitude for the exclusive process $pp \to pp \pi^+ \pi^-$
(with four-momenta $p_{a}+p_{b}\rightarrow p_{1}+p_{2}+p_{3}+p_{4}$)
can be written formally as
\begin{eqnarray}
{\cal {M}}_{pp \to pp\pi\pi}^{full}
(s,y_{3},y_{4},\bp_{1,t},\bp_{2,t},\bp_{m}) &=&
{\cal {M}}_{pp \to pp\pi\pi}^{bare}
(s,y_{3},y_{4},\bp_{1,t},\bp_{2,t},\bp_{m})\nonumber\\
&+&{\cal {M}}_{pp \to pp\pi\pi}^{rescatt}
(s,y_{3},y_{4},\bp_{1,t},\bp_{2,t},\bp_{m})\,,
\nonumber\\
\label{amp_pipi_full}
\end{eqnarray}
where the auxiliary quantity $\bp_{m}=\bp_{3,t}-\bp_{4,t}$.

The "bare" amplitude can be written as
\begin{eqnarray} \nonumber
\mathcal{M}_{pp\to pp\pi\pi}^{bare}&=&
M_{13}(t_1,s_{13})F_{\pi}(t_{a})\frac{1}{t_{a}-m_{\pi}^{2}}F_{\pi}(t_{a})M_{24}(t_2,s_{24})\\
&+&M_{14}(t_1,s_{14})F_{\pi}(t_{b})\frac{1}{t_{b}-m_{\pi}^{2}}F_{\pi}(t_{b})M_{23}(t_2,s_{23})\; ,
\label{Regge_amplitude}
\end{eqnarray}
where $M_{ik}$ denotes "interaction" between nucleon $i=1$ (forward
nucleon) or $i=2$ (backward nucleon) and one of the two pions $k=3$
($\pi^{+}$), $k=4$ ($\pi^{-}$). The energy dependence of the
amplitudes of the $\pi N$ subsystems was parameterized in the Regge
form with pomeron and reggeon exchanges. The details of the matrix
element are explained in Ref.~\cite{LS10}. The strength parameters
and values of the pomeron and reggeon trajectories are taken from
the Donnachie-Landshoff analysis \cite{DL92} of the total cross
section for $\pi N$ scattering. The slope parameters of the elastic
$\pi N$ scattering can be written as shown in Eq.~(\ref{slope_NN})
and are $B_{I\!\!P}^{\pi N}$ = 5.5 GeV$^{-2}$,
$B_{I\!\!R}^{\pi N}$ = 4 GeV$^{-2}$ \cite{LS10}, for pomeron and
reggeon exchanges, respectively.

The Donnachie-Landshoff parametrization \cite{DL92} can be used only
for the $\pi N$ subsystem energy $W_{ik}>2-3$ GeV (see
Ref.~\cite{LS10}). Bellow $W_{ik}=2$ GeV the resonance states in
$\pi N$ subsystems are present. In principle, their contribution
could and should be included explicitly
\footnote{The higher the center-of-mass energy the smaller the relative resonance
contribution.}. In order to exclude resonance regions we shall
"correct" the parametrization (Eq.~(\ref{Regge_amplitude})) multiplying it
by a purely phenomenological smooth cut-off correction factor
\cite{LS10}
\begin{eqnarray}
f_{cont}^{\pi N}(W_{ik})=
\frac{\exp \left( \frac{W_{ik}-W_{0}}{a}\right)}
     {1+\exp \left( \frac{W_{ik}-W_{0}}{a}\right)} \; .
\label{Correction_factor}
\end{eqnarray}
The parameter $W_{0}=2$ GeV gives the position of the cut, and the
parameter $a=0.2$ GeV describes how sharp the cut-off is. For large
energies $f_{cont}^{\pi N}(W_{ik})\approx 1$ and close to
kinematical threshold $f_{cont}^{\pi N}(W_{ik} \simeq
m_{\pi}+m_{N})\approx 0$.

The extra form factors $F(t_{a})$ and $F(t_{b})$ "correct" for
off-shellness of the intermediate pions in the middle of the
diagrams shown in
Fig.~\ref{fig:central_double_diffraction_diagrams}. In the following
they are parameterized as
\begin{equation}
F_{\pi}(t_{a,b})=\exp\left(\frac{t_{a,b}-m_{\pi}^{2}}{\Lambda^{2}_{off, E}}\right) \;,
\label{off-shell_form_factors}
\end{equation}
i.e. normalized to unity on the pion-mass-shell. In general, the
parameter $\Lambda_{off, E}$ is not known precisely but, in
principle, could be fitted to the (normalized) experimental data.
From our general experience in hadronic physics we expect
$\Lambda_{off, E}^{2} \sim 1.2 - 2$ GeV$^{2}$. How to extract
$\Lambda_{off, E}$ will be discussed in the result section.

The absorptive corrections to the "bare" amplitude
(Eq.~(\ref{Regge_amplitude})) can be written as:
\begin{eqnarray}
{\cal M}_{pp\to pp\pi\pi}^{rescat}=
\mathrm{i}
\int \frac{d^{2} \bk_{t}}{2(2\pi)^{2}} \frac{A_{pp}(s,k_{t}^{2})}{s}
{\cal M}_{pp\to pp\pi\pi}^{bare}
(\bp^{\,*}_{a,t}-\bp_{1,t},\bp^{\,*}_{b,t}-\bp_{2,t}) \;,
\label{abs_correction}
\end{eqnarray}
where $p^{\,*}_{a} = p_{a} - k_{t}$, $p^{\,*}_{b} = p_{b} + k_{t}$.

The amplitude described above (see Eq.~\ref{amp_pipi_full}) is used to calculate
the corresponding cross section including limitations of the
four-body phase-space. The cross section for the two-pion continuum
is obtained by integration over the four-body phase space:
\begin{eqnarray}
\sigma =\int \frac{1}{2s} \overline{ |{\cal M}|^2} (2 \pi)^4
\delta^4 (p_a + p_b - p_1 - p_2 - p_3 - p_4)
\frac{d^3 p_1}{(2 \pi)^3 2 E_1}
\frac{d^3 p_2}{(2 \pi)^3 2 E_2}
\frac{d^3 p_3}{(2 \pi)^3 2 E_3}
\frac{d^3 p_4}{(2 \pi)^3 2 E_4}. \nonumber \\
\;
\label{dsigma_for_2to4}
\end{eqnarray}
In order to calculate the total cross section one has to calculate
the eight-dimensional integral numerically. The details how to
conveniently reduce the number of kinematical integration variables
are given elsewhere \cite{LS10}.

%--------------------
\section{Results}
\label{section:III}
%--------------------

Let us start from presenting various differential cross sections. In
practical integrations of the exclusive $\chi_{c0}$ meson we choose
the transverse momenta of outgoing nucleons $(p_{1,t}, p_{2,t})$,
the meson rapidity $(y)$ and the relative azimuthal angle between
outgoing nucleons $(\phi_{12})$.

\begin{figure}[!h]
\includegraphics[width = 0.32\textwidth]{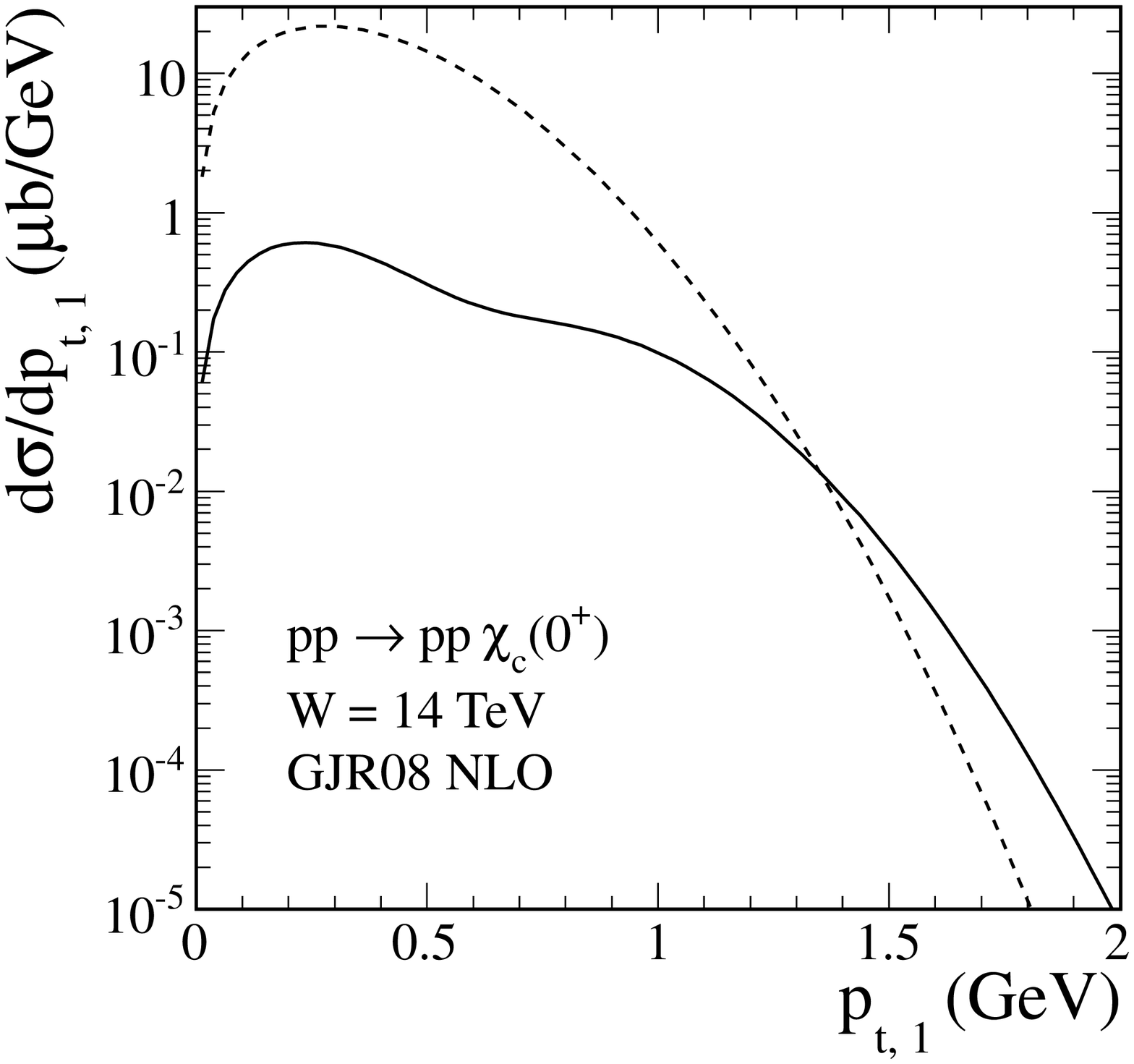}
\includegraphics[width = 0.32\textwidth]{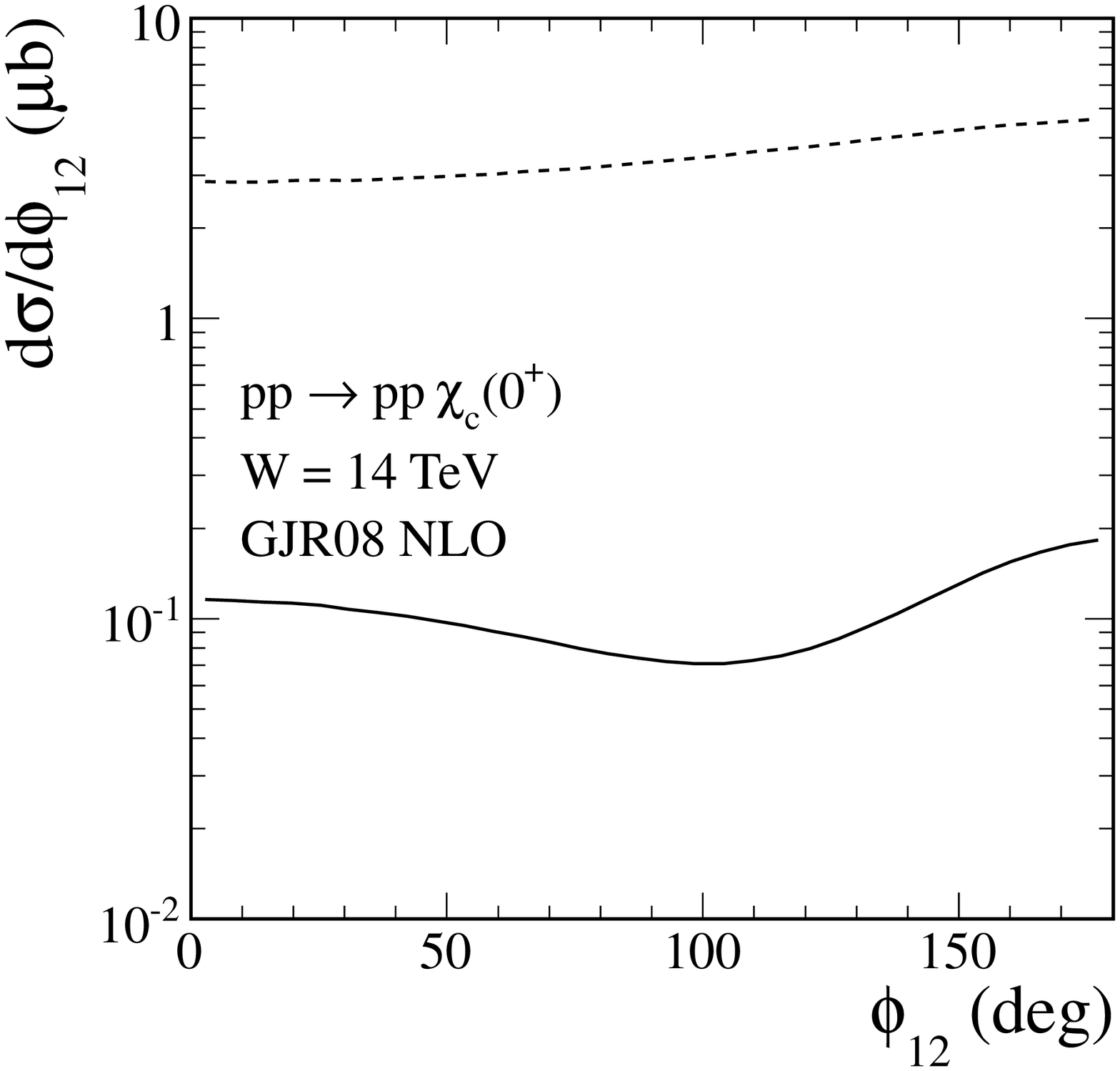}\\
\includegraphics[width = 0.32\textwidth]{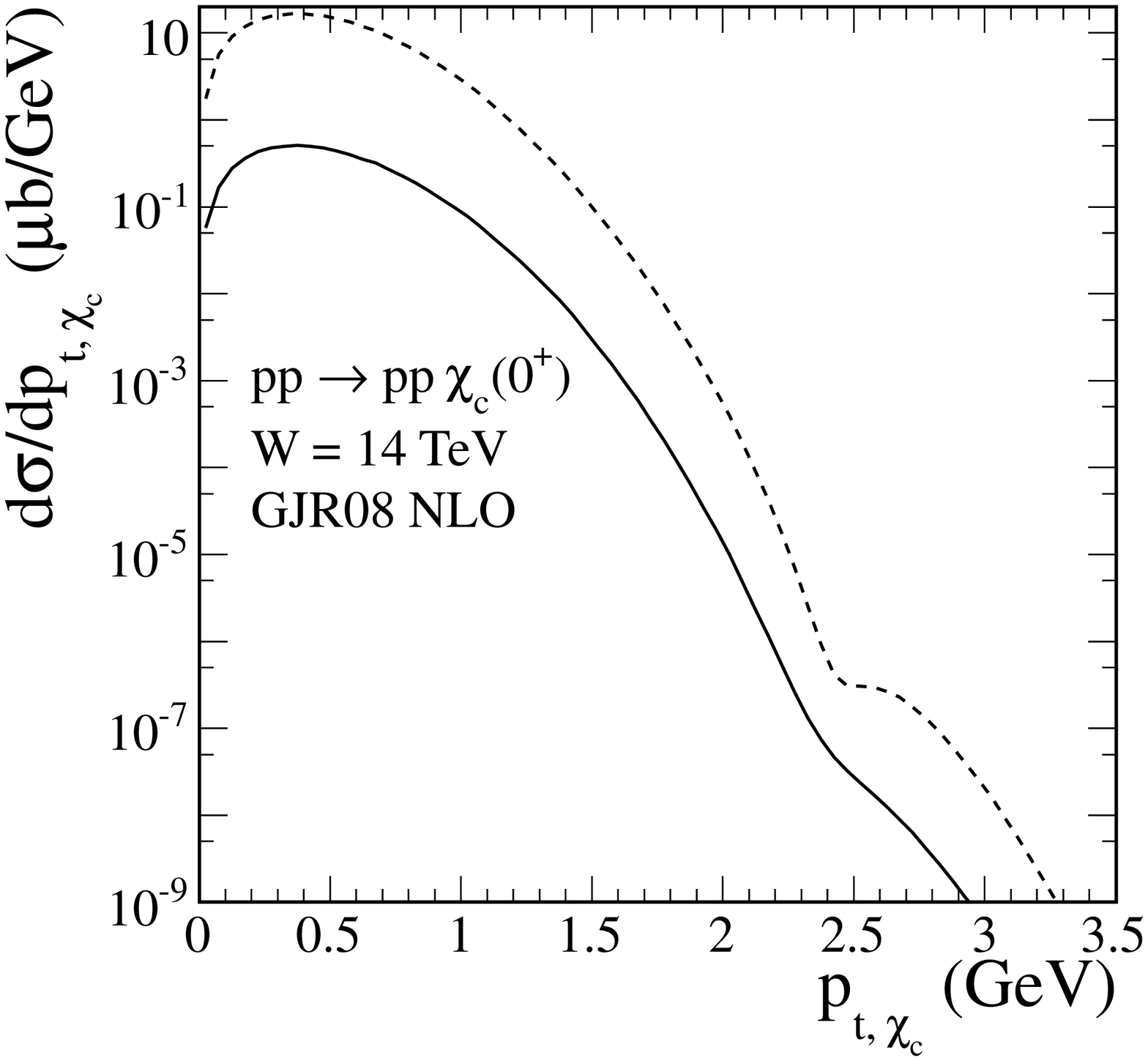}
\includegraphics[width = 0.32\textwidth]{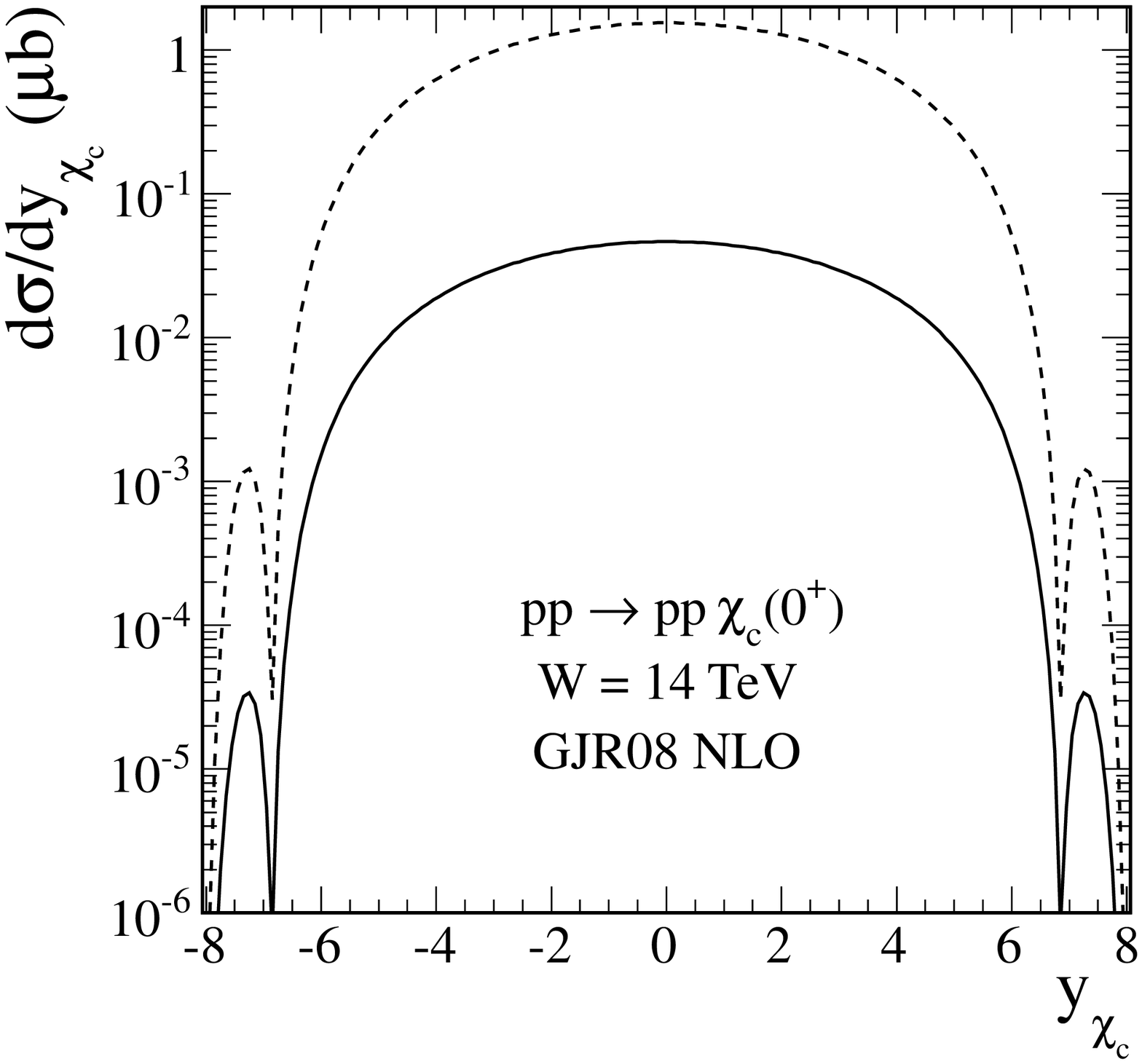}
  \caption{\label{fig:diff_comp}
  \small
Differential cross sections for the $pp \to pp \chi_{c0}$ reaction
at $\sqrt{s}$ = 14 TeV without (dashed line) and with (solid line)
absorption effects. These calculations were done with the GJR08 NLO
\cite{GJR} UGDFs.}
\end{figure}

In Fig.~\ref{fig:diff_comp} we show distributions of the central
exclusive $\chi_{c0}$ production cross section at $\sqrt{s}$ = 14
TeV without (dashed line) and with (solid line) absorptive
corrections as described in subsection \ref{subsection:signal}.
These calculations were done with GJR NLO \cite{GJR} collinear gluon
distribution, to generate the KMR UGDFs (see Eq.~(\ref{fg_kmr})),
which allows to use low values of the gluon transverse momenta
$Q_{t}^2 \geq Q_{cut}^2 = 0.5$ GeV$^2$. The bigger the value of the
cut-off parameter, the smaller the cross section (see
Ref.~\cite{PST_chic0}). In the calculations we take the value of the
hard scale to be $\mu^2 = M^{2}/4$. The smaller $\mu^2$, the bigger
the cross section \cite{PST_chic0}. 
The absorption effects lead to a damping of the cross section.
In most distributions the shape is almost unchanged.
Exception is the distribution in proton transverse momentum where
the absorption effects lead to a damping of the cross section at small
proton $p_{t}$ and an enhancement of the cross section at large proton $p_{t}$. 
In relative azimuthal angle distribution we observe a "diffractive dip" structure 
in the region of
$\phi_{12}\sim \pi/2$. Transverse momentum distribution of
$\chi_{c0}$ shows a small minimum at $p_t \sim$ 2.5 GeV. The main reason of its
appearance is the functional dependence of matrix elements on its
arguments \cite{PST_chic0}.
%--------------------------------------------------------
%--------------------------------------------------------

\begin{figure}[!h]
\includegraphics[width = 0.32\textwidth]{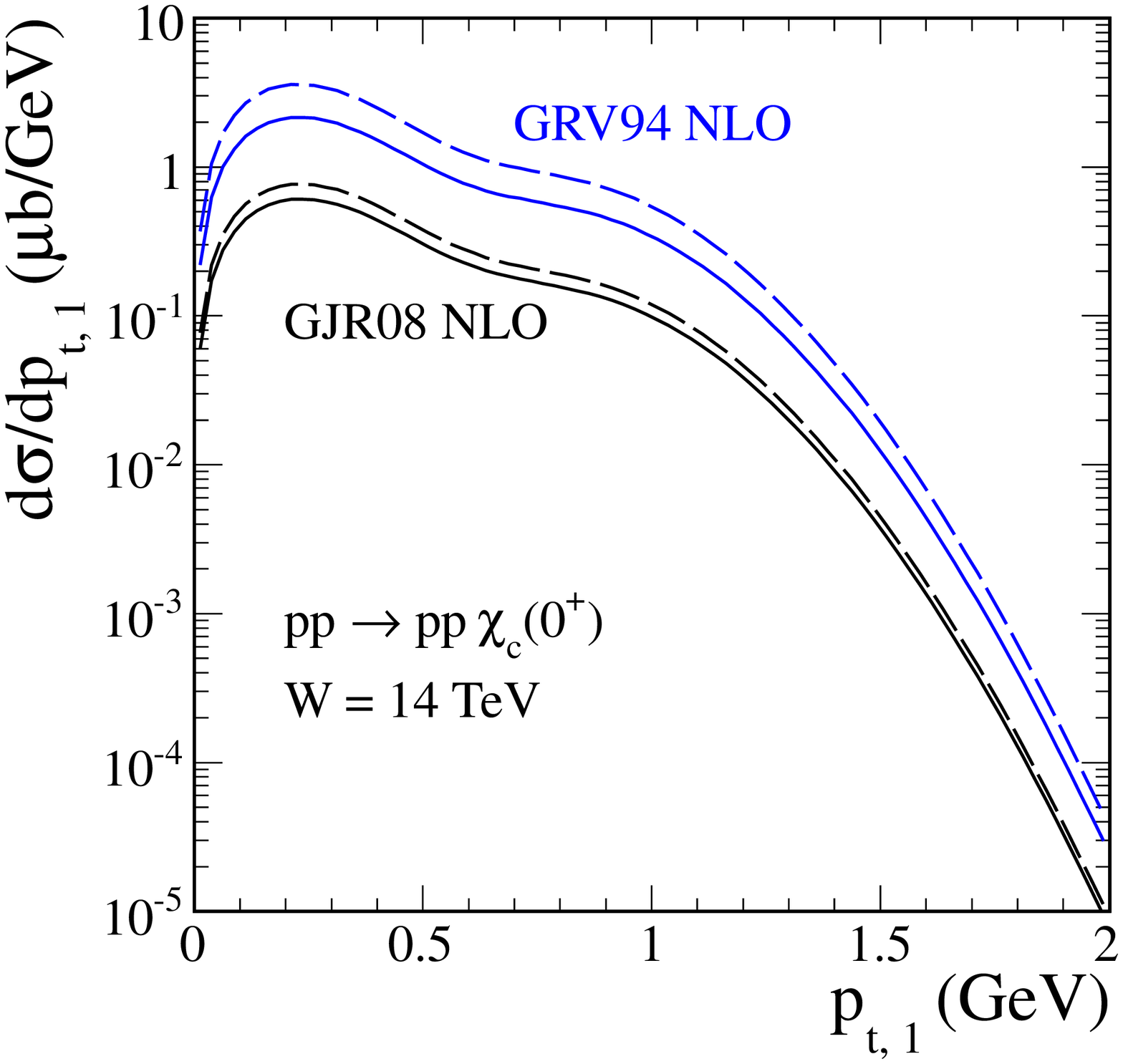}
\includegraphics[width = 0.32\textwidth]{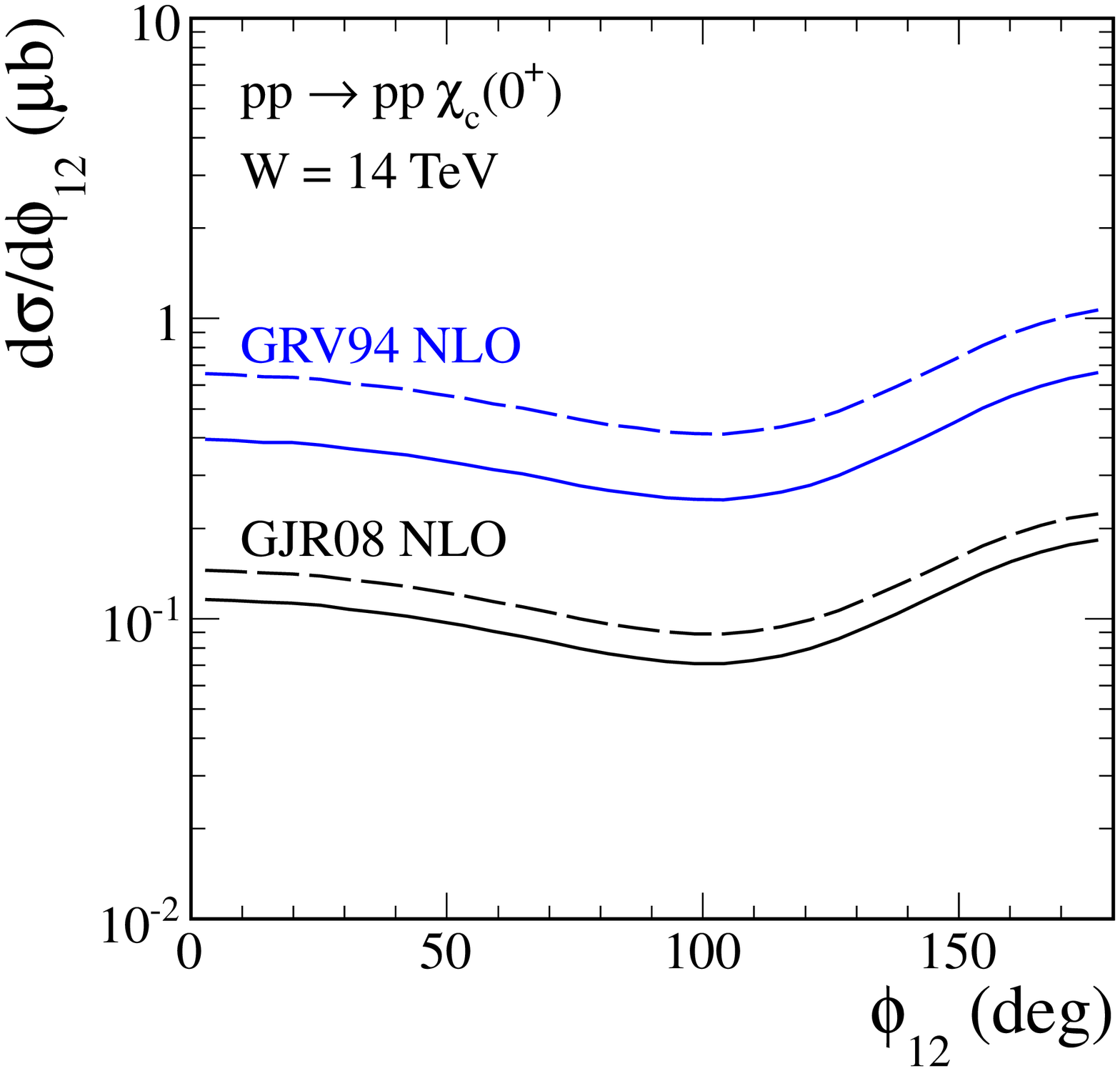}\\
\includegraphics[width = 0.32\textwidth]{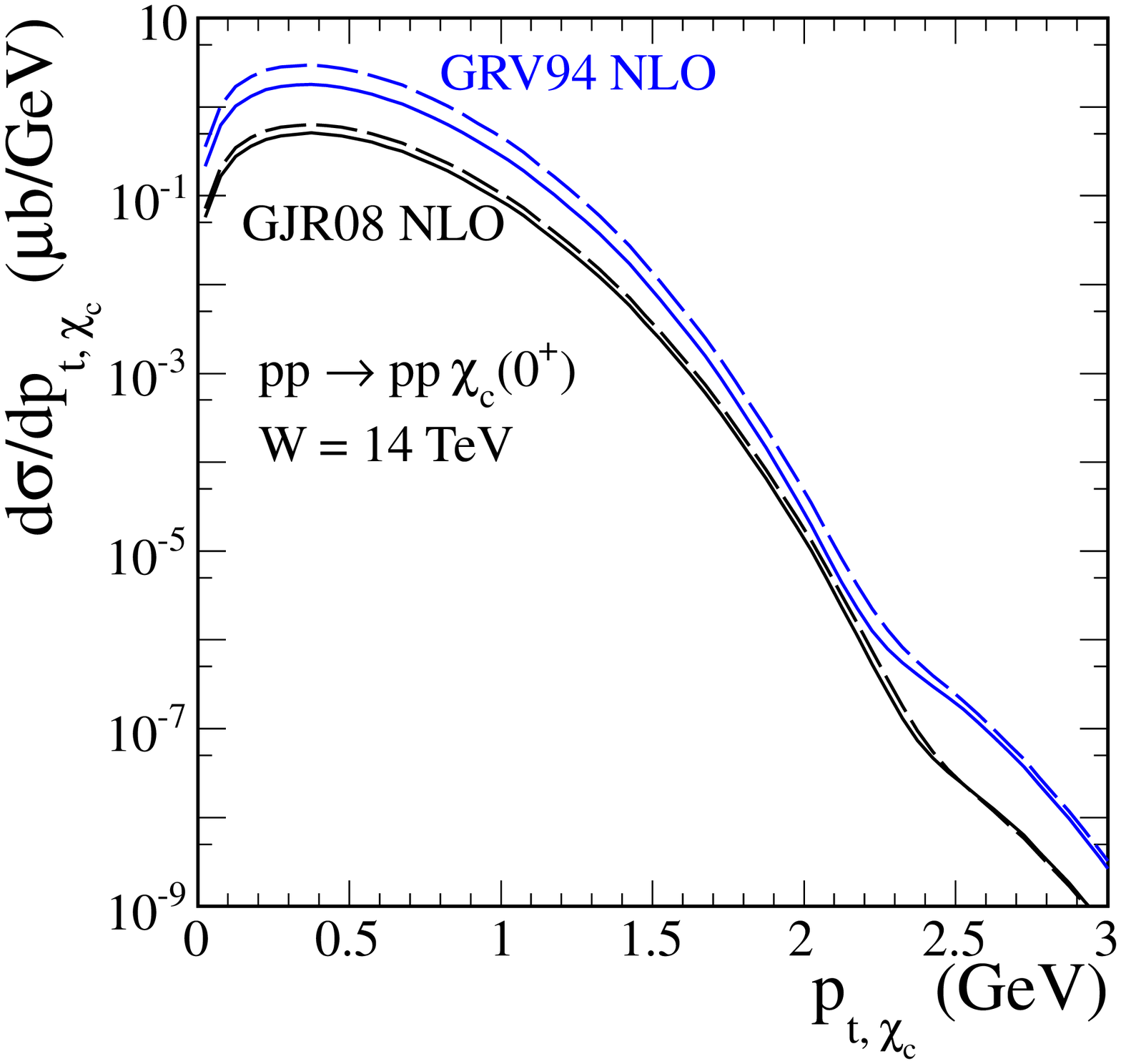}
\includegraphics[width = 0.32\textwidth]{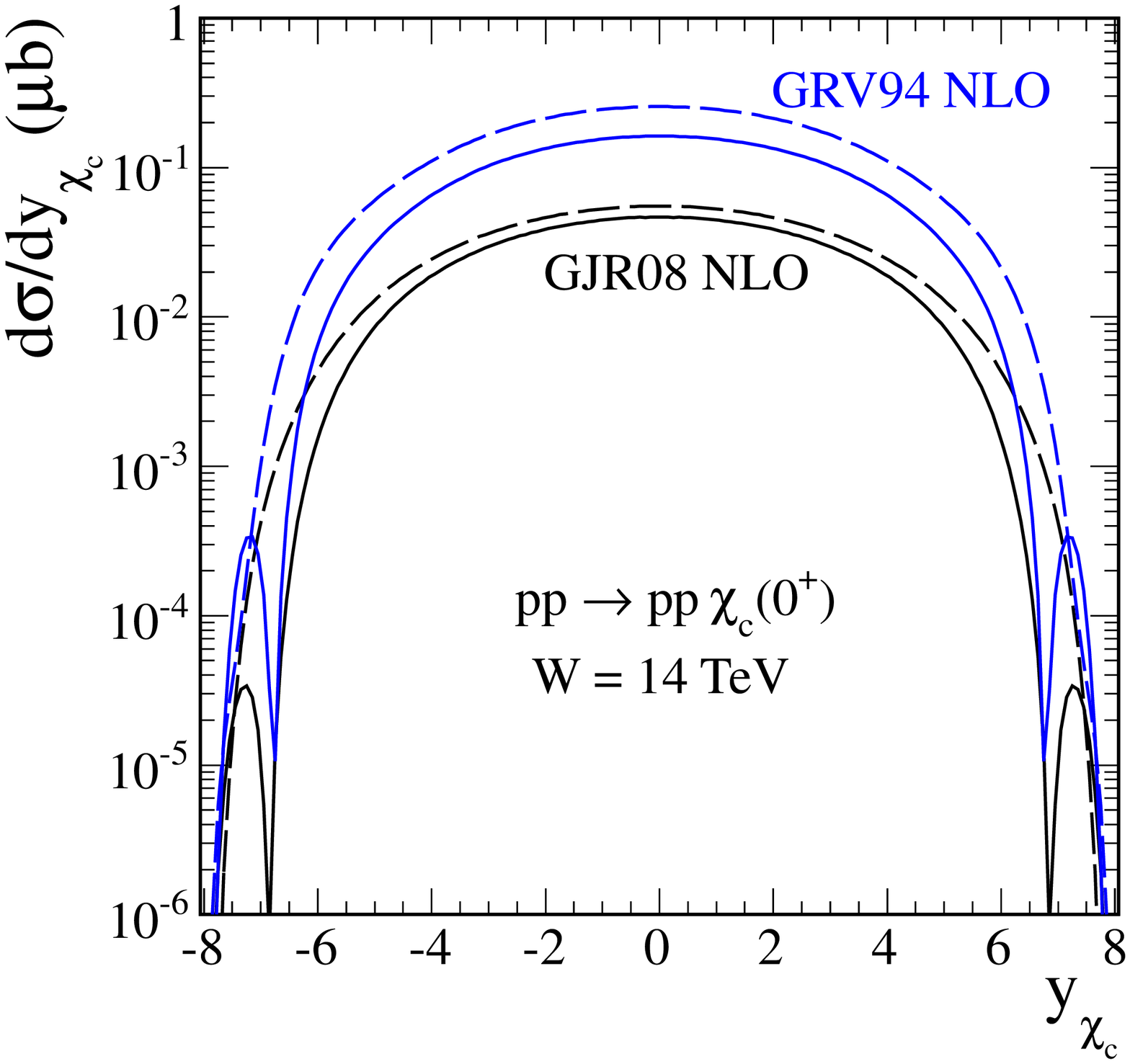}
  \caption{\label{fig:diff_ugdfs}
  \small
Differential cross sections for the $pp \to pp \chi_{c0}$ reaction
at $\sqrt{s}$ = 14 TeV with absorption effects.
The results with KMR off-diagonal UGDFs given by Eq.~(\ref{fg_kmr}) (solid lines) 
and with off-diagonal UGDFs given by Eq.~(\ref{phenom_form}) (dashed lines) are shown.
These calculations were done for two UGDFs:
GRV94 NLO (upper lines) and GJR08 NLO (bottom lines).
}
\end{figure}

In Fig.~\ref{fig:diff_ugdfs} we compare distributions of the central
exclusive $\chi_{c0}$ production cross section at $\sqrt{s}$ = 14
TeV calculated for two collinear gluon distributions: GRV94 NLO
(upper lines) and GJR08 NLO (bottom lines). We show results here
only for distributions with absorptive corrections calculated with
the KMR off-diagonal UGDFs given by Eq.~(\ref{fg_kmr}) (solid lines) and
with off-diagonal UGDFs given in the
phenomenological form given by Eq.~(\ref{phenom_form}).
The peaks at large rapidities appear only when we use formula (\ref{fg_kmr}).
In this region one of off-diagonal UGDFs changes a sign.
This shows limitations in applying formula (\ref{fg_kmr}).
%--------------------------------------------------------
%--------------------------------------------------------

\begin{figure}[!h]
\includegraphics[width = 0.4\textwidth]{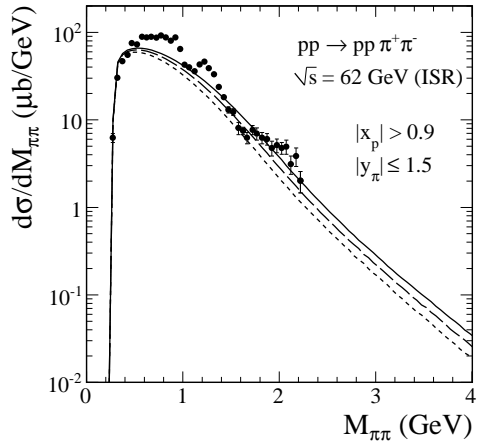}
  \caption{\label{fig:dsig_dmpipi_ISR}
  \small
Differential cross section $d\sigma/dM_{\pi\pi}$ 
for the $pp \to pp \pi^{+} \pi^{-}$ reaction 
at $\sqrt{s} = 62$ GeV with experimental
cuts relevant for the CERN ISR experimental data from Ref.~\cite{ABCDHW90}.
% (left panel) and data from Ref.~\cite{WSW83} (right panel).
Results with the cut-off parameters $\Lambda_{off,E}^{2}=1.6, 1.8,
2.0$ GeV$^{2}$ are shown by the dotted, dashed and solid line,
respectively. The absorption effects were included in the calculations.}
\end{figure}

Let us turn now to an estimation of the two-pion background. In
Fig.~\ref{fig:dsig_dmpipi_ISR} we show the two-pion invariant mass
distribution at the center-of-mass energy of the CERN ISR
$\sqrt{s} = 62$ GeV
% (left panel) and $\sqrt{s} = 63$ GeV (right panel)
(this is the highest energy at which experimental data exist).
Here we have used a simple model described in subsection
\ref{subsection:baskground}. The experimental cuts on the rapidity
of both pions and on longitudinal momentum fractions
(Feynman-$x$, $x_{F} = 2p_{\parallel}/\sqrt{s}$)
of both outgoing protons are included when comparing our results
with existing experimental data from \cite{ABCDHW90}
(see Refs.~\cite{WSW83,ABCDHW89} for early studies).
%and data from \cite{WSW83} (right panel).
Absorption effects were included in this calculation. 
The results depend on the value of
the nonperturbative, a priori unknown parameter of the form factor
responsible for off-shell effects (see
Eq.~(\ref{off-shell_form_factors})). We show results with the
cut-off parameters $\Lambda_{off,E}^{2}=1.6, 1.8, 2.0$ GeV$^{2}$ as
represented by the dotted, dashed and solid line, respectively. The
experimental data show some peaks above our flat model continuum.
They correspond to the well known $\pi^+ \pi^-$ resonances:
$\sigma$, $\rho^0$, $f_2(1275)$ which are not included explicitly in
our calculation. In the present analysis we are interested mostly
what happens above $M_{\pi \pi} > 3$ GeV, above the resonance region
where there are no experimental data points. Our model with
$\Lambda_{off,E}$ parameter fitted to the data provides an educated
extrapolation to the unmeasured region.
%--------------------------------------------------------
%--------------------------------------------------------

\begin{figure}[!h]
\includegraphics[width = 0.4\textwidth]{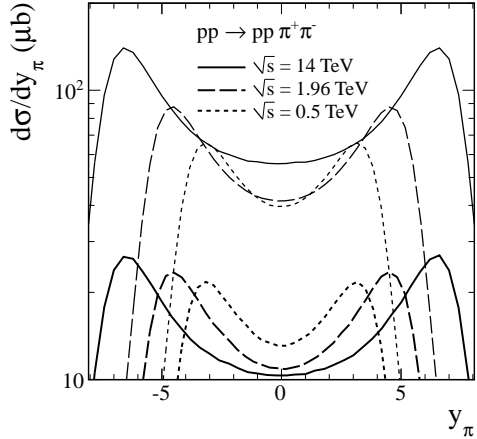}
  \caption{\label{fig:y_pipi_comp}
  \small
Differential cross section $d\sigma/dy_{\pi}$
for the $pp \to pp \pi^{+} \pi^{-}$ reaction
at $\sqrt{s} = 0.5, 1.96, 14$ TeV
with $\Lambda_{off,E}^{2}=2.0$ GeV$^{2}$.
The results without (upper lines) and with (bottom lines)
absorption effects are shown.
}
\end{figure}

In Fig.~\ref{fig:y_pipi_comp} we show differential distributions in
pion rapidity for the $pp \to pp \pi^{+} \pi^{-}$ reaction at
$\sqrt{s} = 0.5, 1.96, 14$ TeV without (upper lines) and with
(bottom lines) absorption effects. The integrated cross section
slowly rises with incident energy. The camel-like shape of the
distributions is due to the interference of components in the
amplitude: pomeron-pomeron component dominates at midrapidities of pions
and pomeron-reggeon (reggeon-pomeron) peaks at backward (forward) pion
rapidities, respectively (see Ref.~\cite{LS10}).
%The absorption effects reverse the trend at midrapidity of pions.
%In general, the higher energy the higher absorption effects.
The reader is asked to notice that
the energy dependence of the cross section at $y_{\pi} \approx 0$
is reversed by the absorption effects which
are stronger at higher energies.
%--------------------------------------------------------
%--------------------------------------------------------

\begin{figure}[!h]
\includegraphics[width = 0.32\textwidth]{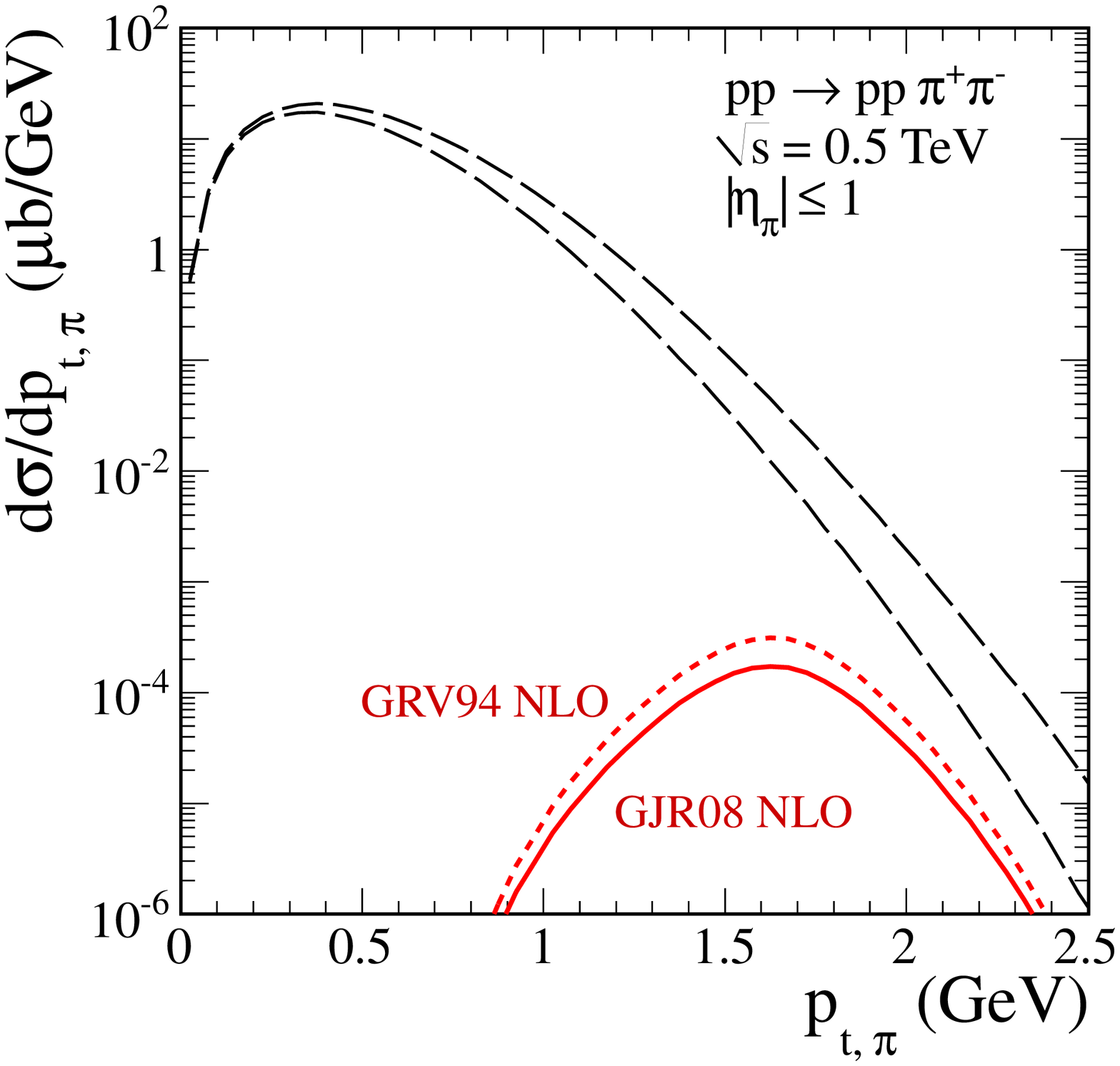}
\includegraphics[width = 0.32\textwidth]{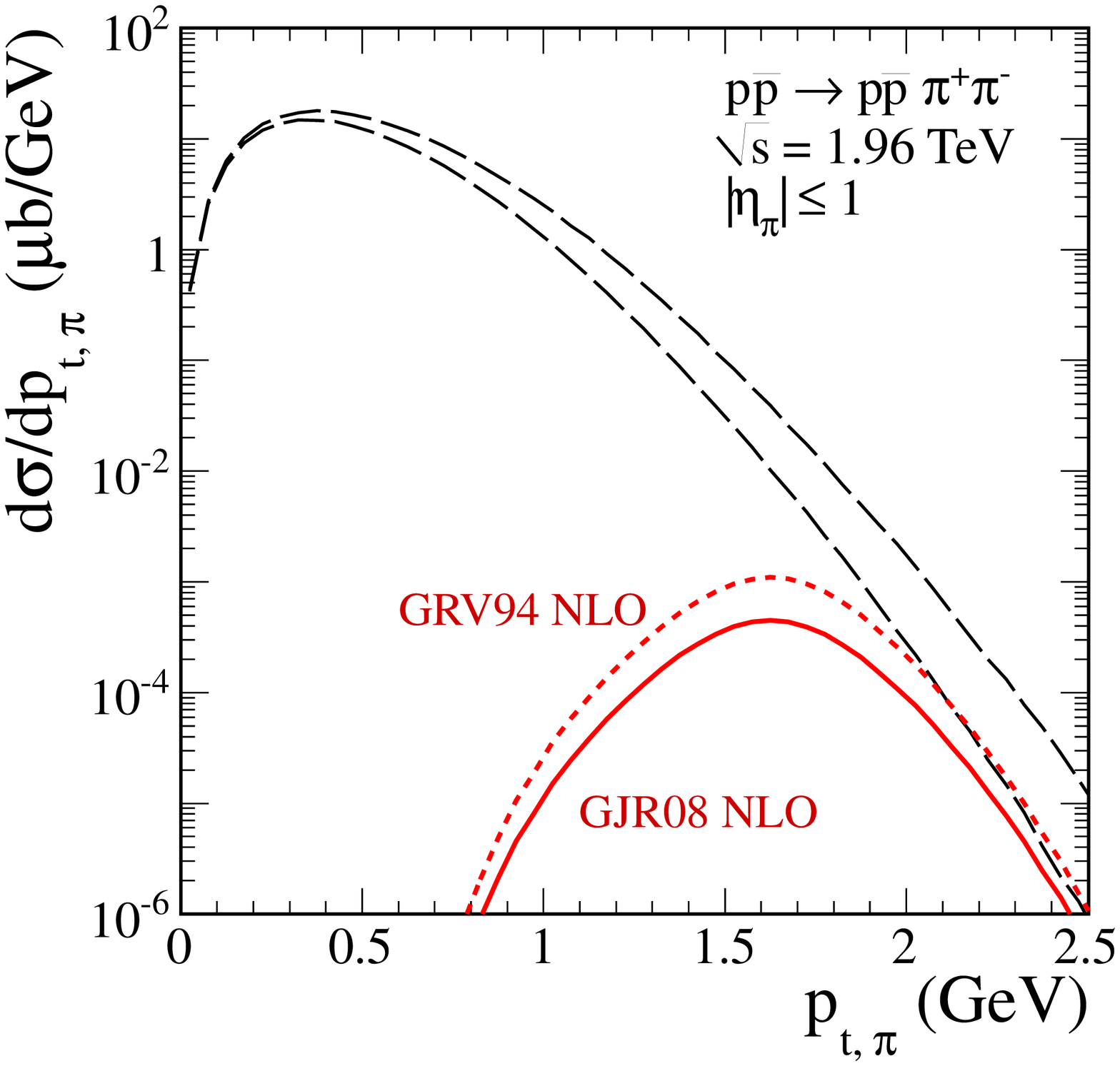}
\includegraphics[width = 0.32\textwidth]{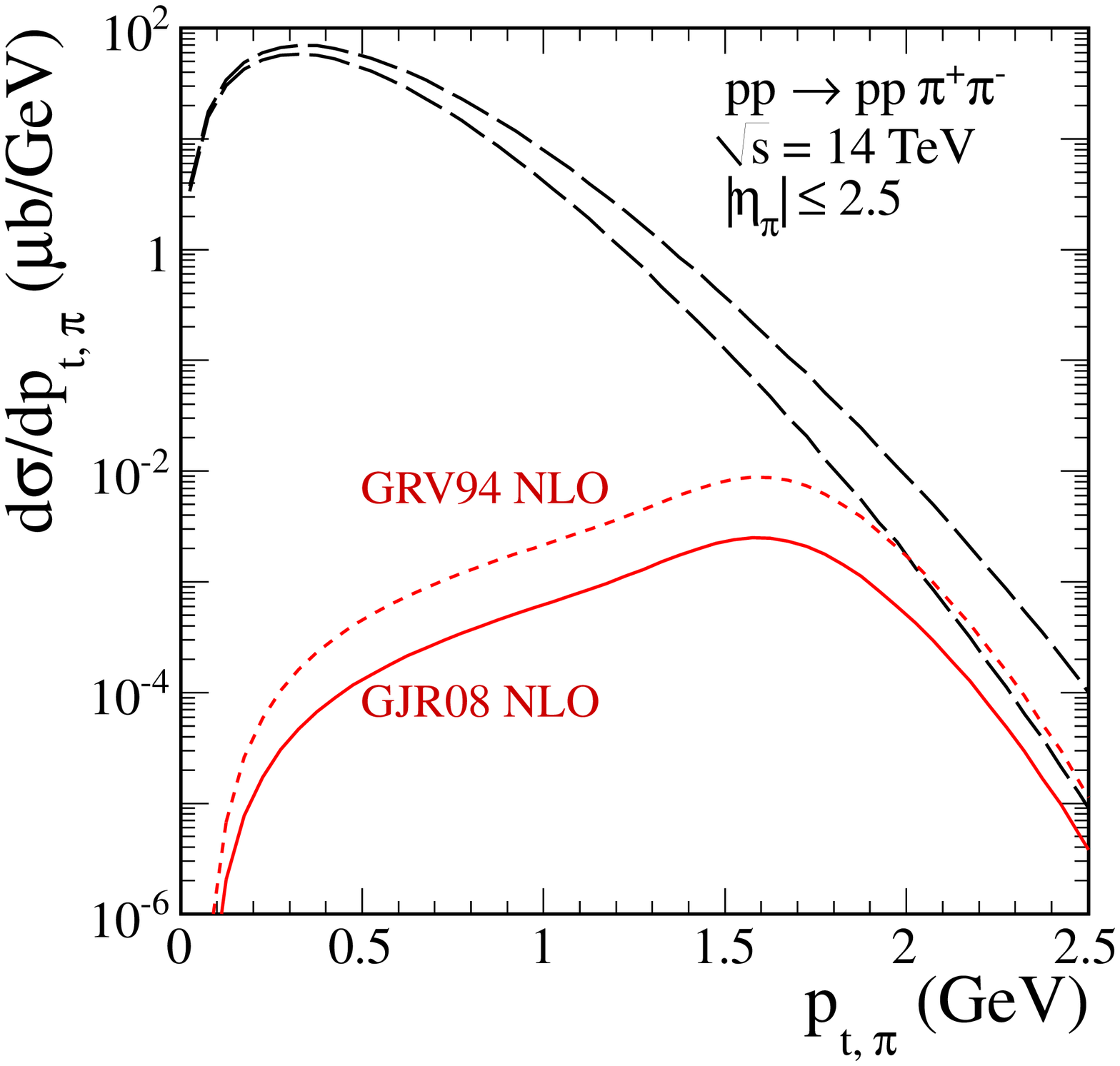}
  \caption{\label{fig:pt_lhc}
  \small
Differential cross section $d\sigma/dp_{t,\pi}$ at $\sqrt{s} = 0.5,
1.96, 14$ TeV with cuts on the pion pseudorapidities. The
diffractive background was obtained with the cut-off parameters
$\Lambda_{off, E}^{2}$ = 1.6, 2.0 GeV$^{2}$ (lower and upper dashed
lines, respectively). Results for the pions from the decay of the
$\chi_{c0}$ meson including the $\pi^{+}\pi^{-}$ branching ratio, for GRV94 NLO
(upper lines) and GJR08 NLO (bottom lines) UGDFs, are shown. The
absorption effects were included in the calculations.}
\end{figure}

Now we wish to compare differential distributions of pions from the
$\chi_{c0}$ decay with those for the continuum pions. In the first
step we calculate the two-dimensional distribution
$d\sigma(y,p_{t})/dy dp_{t}$, where $y$ is rapidity and $p_{t}$ is the transverse
momentum of $\chi_{c}$. The decay of $\chi_{c0} \to \pi^{+}\pi^{-}$
is included then in a simple Monte Carlo program assuming isotropic
decay of the scalar $\chi_{c0}$ meson in its rest frame. The
kinematical variables of pions are transformed to the overall
center-of-mass frame where extra cuts are imposed. Including the
simple cuts we construct several differential distributions in
different kinematical variables. In Fig.~\ref{fig:pt_lhc} we show
distributions in pion transverse momenta. The pions from the decay
are placed at slightly larger transverse momenta. This can be
therefore used to get rid of the bulk of the continuum by imposing
an extra cut on the pion transverse momenta.
%--------------------------------------------------------
%--------------------------------------------------------

\begin{figure}[!h]
\includegraphics[width = 0.32\textwidth]{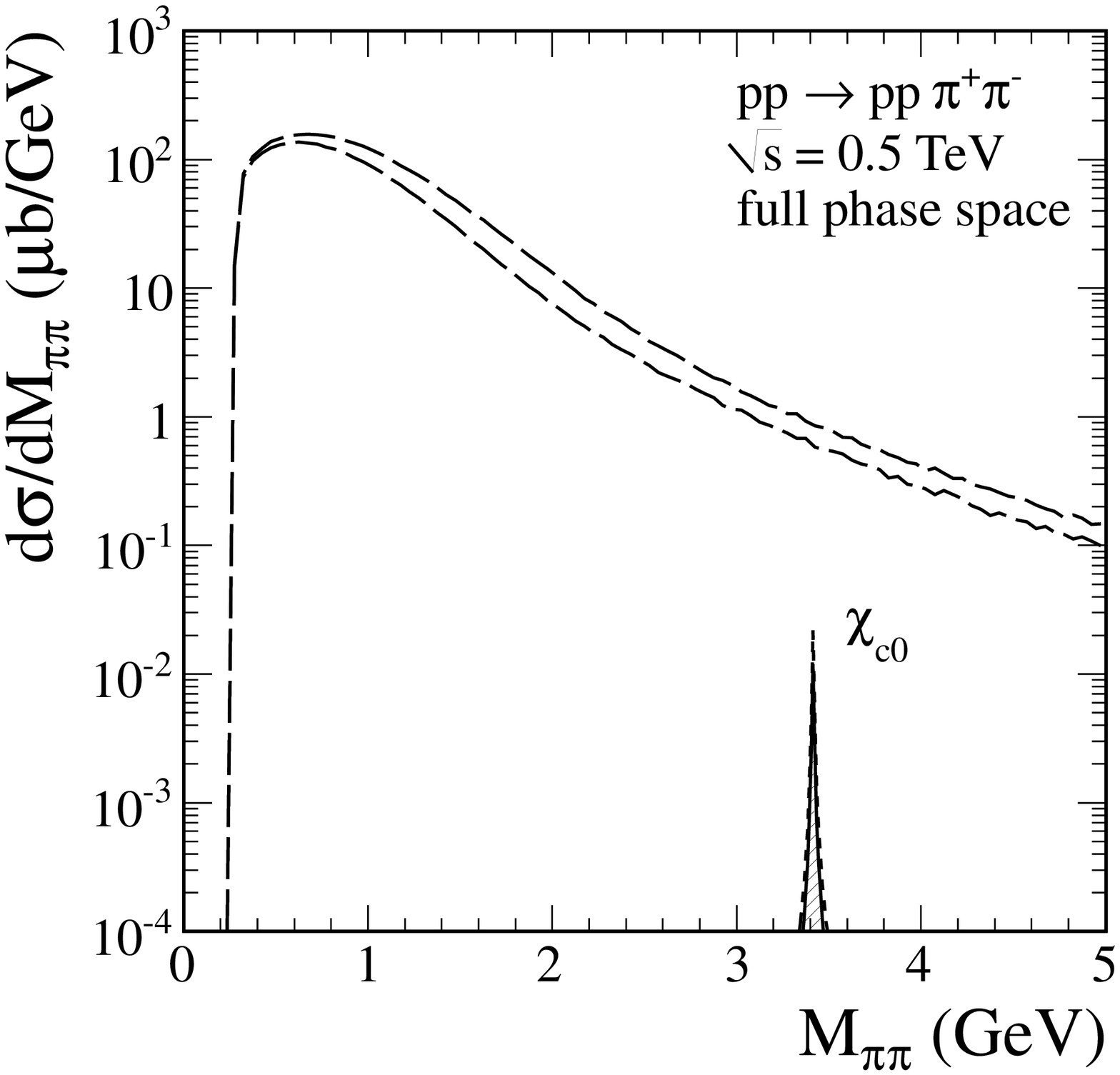}
\includegraphics[width = 0.32\textwidth]{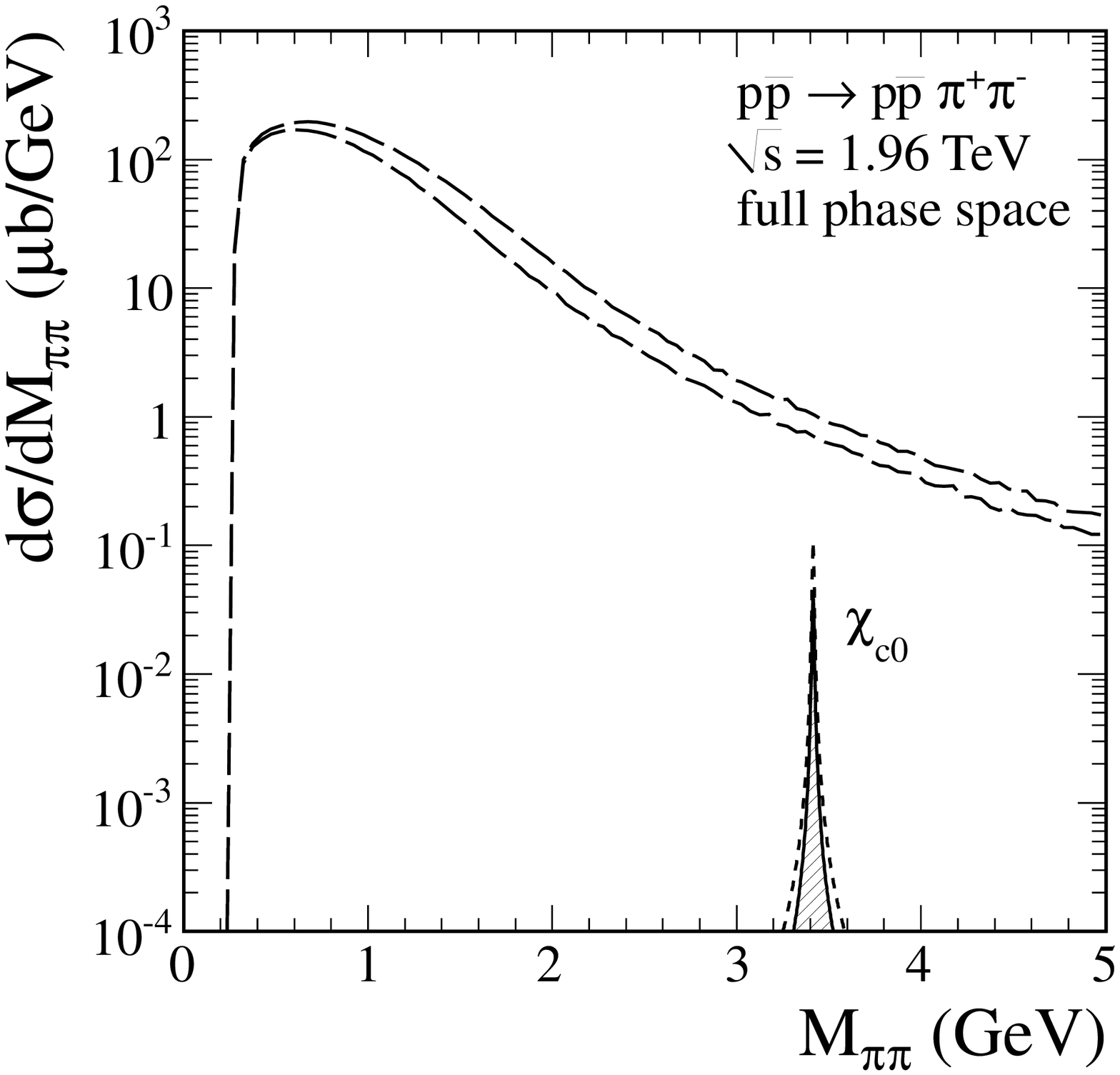}
\includegraphics[width = 0.32\textwidth]{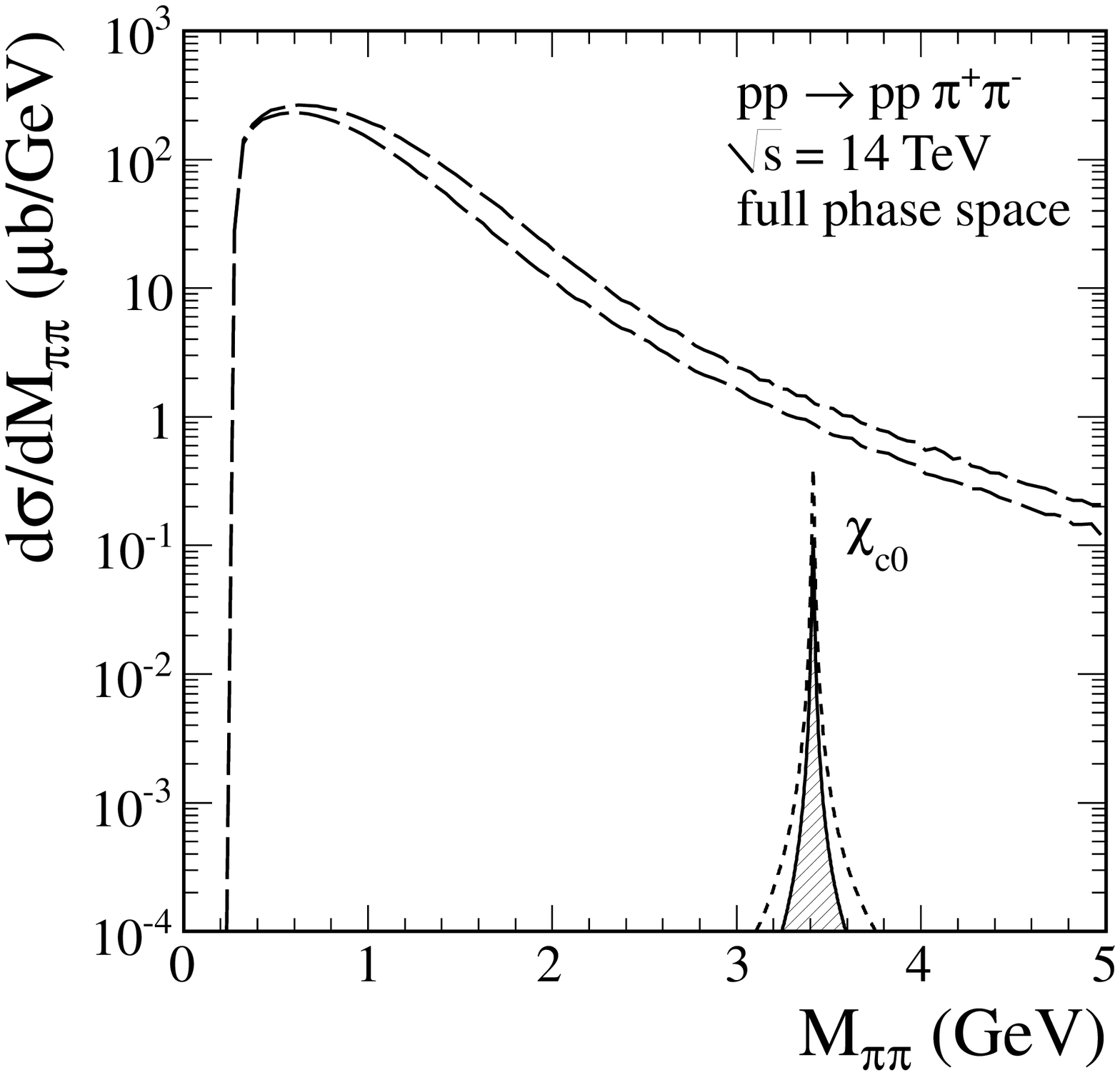}\\
\includegraphics[width = 0.32\textwidth]{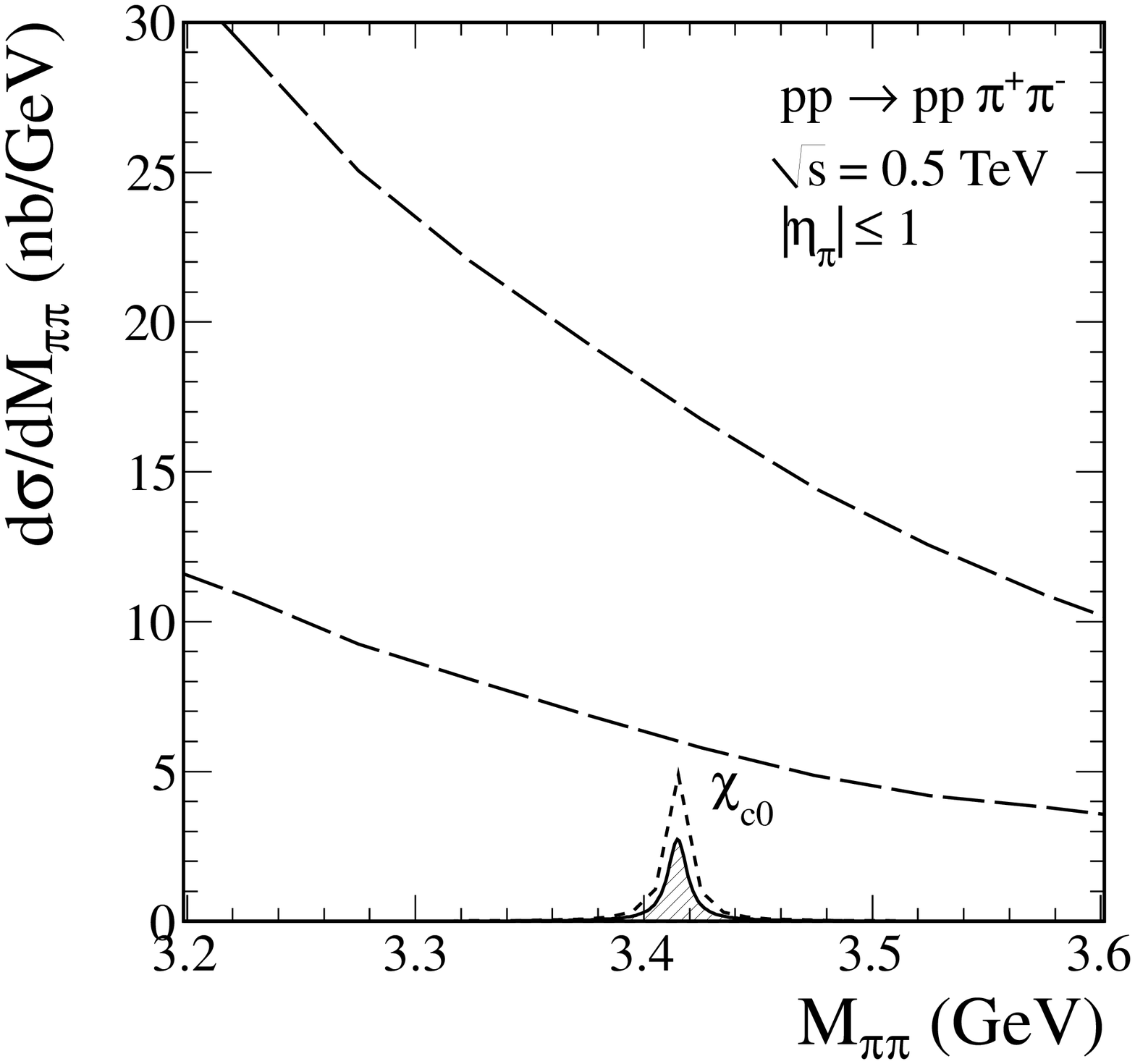}
\includegraphics[width = 0.32\textwidth]{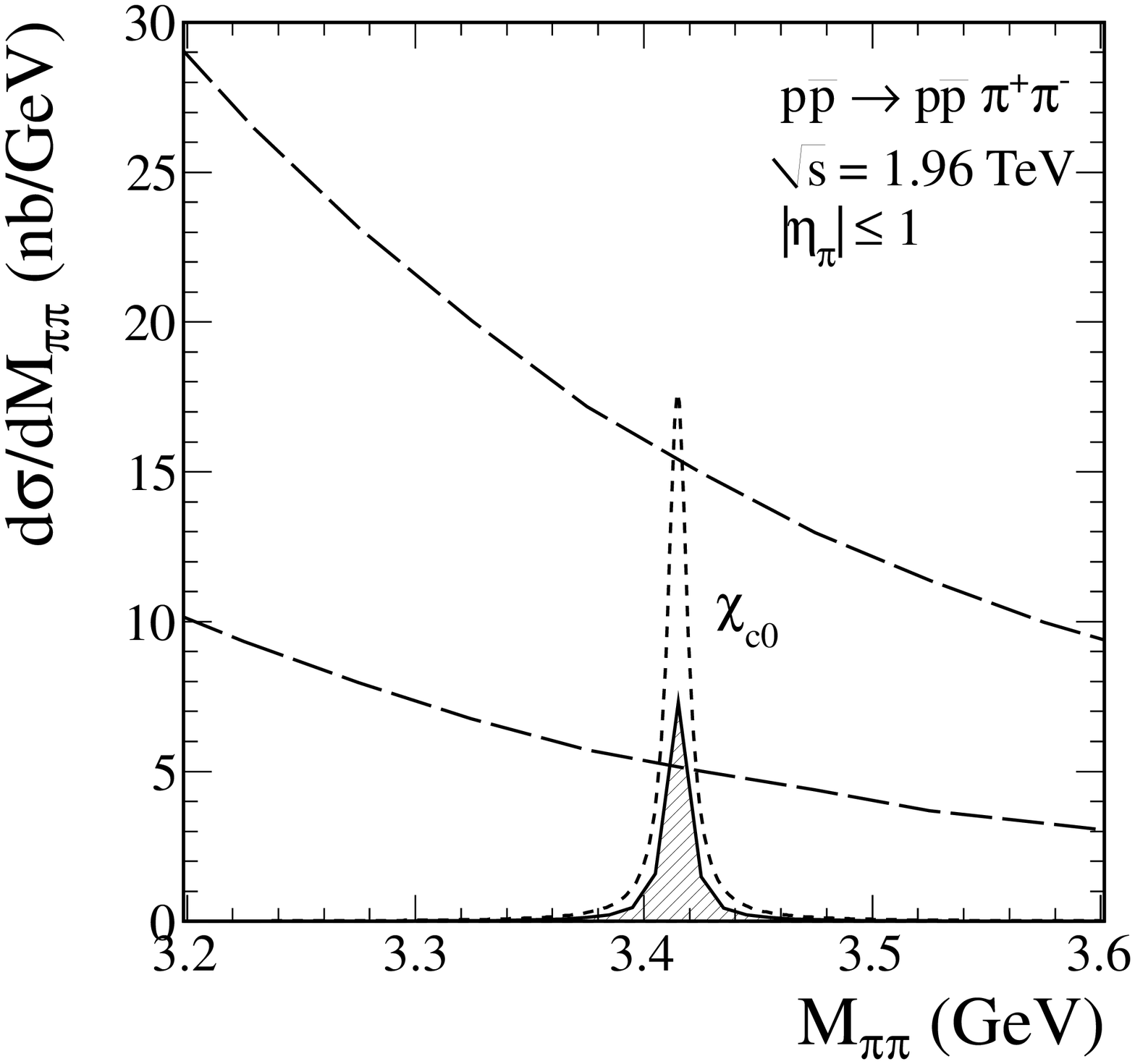}
\includegraphics[width = 0.32\textwidth]{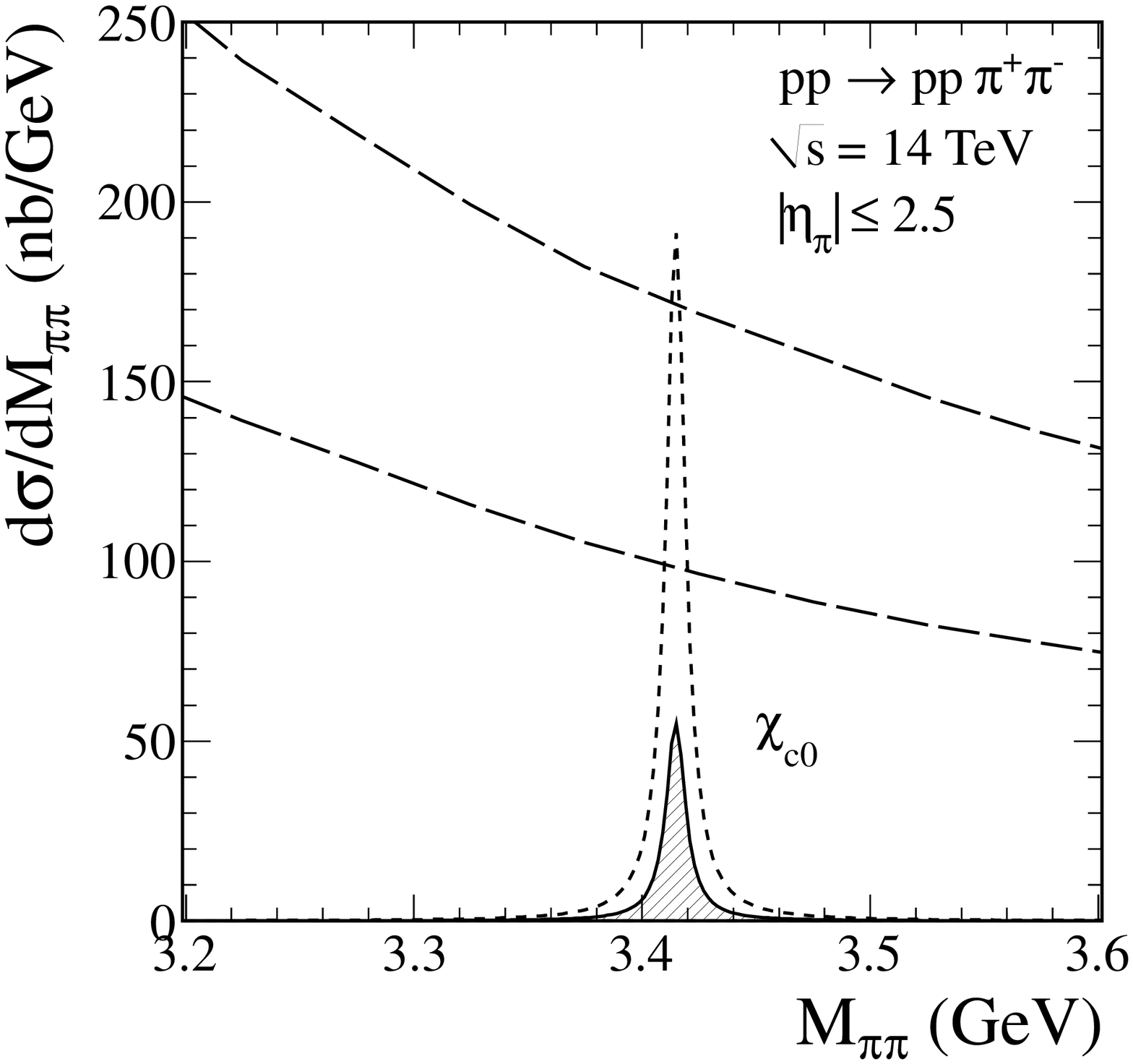}
  \caption{\label{fig:dsig_dmpipi}
  \small
The $\pi^{+}\pi^{-}$ invariant mass distribution at $\sqrt{s} = 0.5,
1.96, 14$ TeV integrated over the full phase space (upper row) and
with the detector limitations in pion pseudorapidities (lower row).
The dashed lines present the $\pi\pi$ continuum with the cut-off
parameters $\Lambda_{off, E}^{2}$ = 1.6, 2.0 GeV$^{2}$ (lower and
upper dashed lines, respectively). The $\chi_{c0}$ contribution
is calculated with GRV94 NLO (dotted line) and GJR08 NLO (solid line) collinear
gluon distributions. The absorption effects for the $\chi_{c0}$
meson and for the background were included in the calculations.}
\end{figure}

In Fig.~\ref{fig:dsig_dmpipi} we show two-pion invariant mass
distribution for the double-diffractive $\pi\pi$ continuum and the
contribution from the decay of the $\chi_{c0}$ meson (see the peak
at $M_{\pi\pi} \simeq 3.4$ GeV). In these figures the resonant
$\chi_{c0}$ distribution was parameterized in the non-relativistic
Breit-Wigner form:
\begin{eqnarray}
\frac{d\sigma}{dM_{\pi\pi}}=
\mathcal{B}(\chi_{c0} \to \pi^{+}\pi^{-})\, \sigma_{pp \to pp \chi_{c0}}\, 2 M_{\pi\pi} \,
\frac{1}{\pi}
\frac{M_{\pi \pi} \Gamma}
     {(M_{\pi\pi}^{2}-M^{2})^{2} + (M_{\pi\pi} \Gamma)^{2}}\,,
\label{BW_form}
\end{eqnarray}
with parameters according to PDG \cite{PDG}.
In calculation of the $\chi_{c0}$ distribution we use GRV94 NLO and
GJR08 NLO collinear gluon distributions. The cross sections for the
$\chi_{c0}$ production and for the background include absorption
effects. While upper row shows the cross section integrated over the
full phase space at different energies, the lower row shows results
including the relevant pion pseudorapidities restrictions 
$-1 < \eta_{\pi^{+}},\eta_{\pi^{-}} < 1$ (RHIC and Tevatron) and 
$-2.5 < \eta_{\pi^{+}},\eta_{\pi^{-}} < 2.5$ (LHC).

%--------------------------------------------------------
%--------------------------------------------------------

\begin{figure}[!h]
\includegraphics[width = 0.32\textwidth]{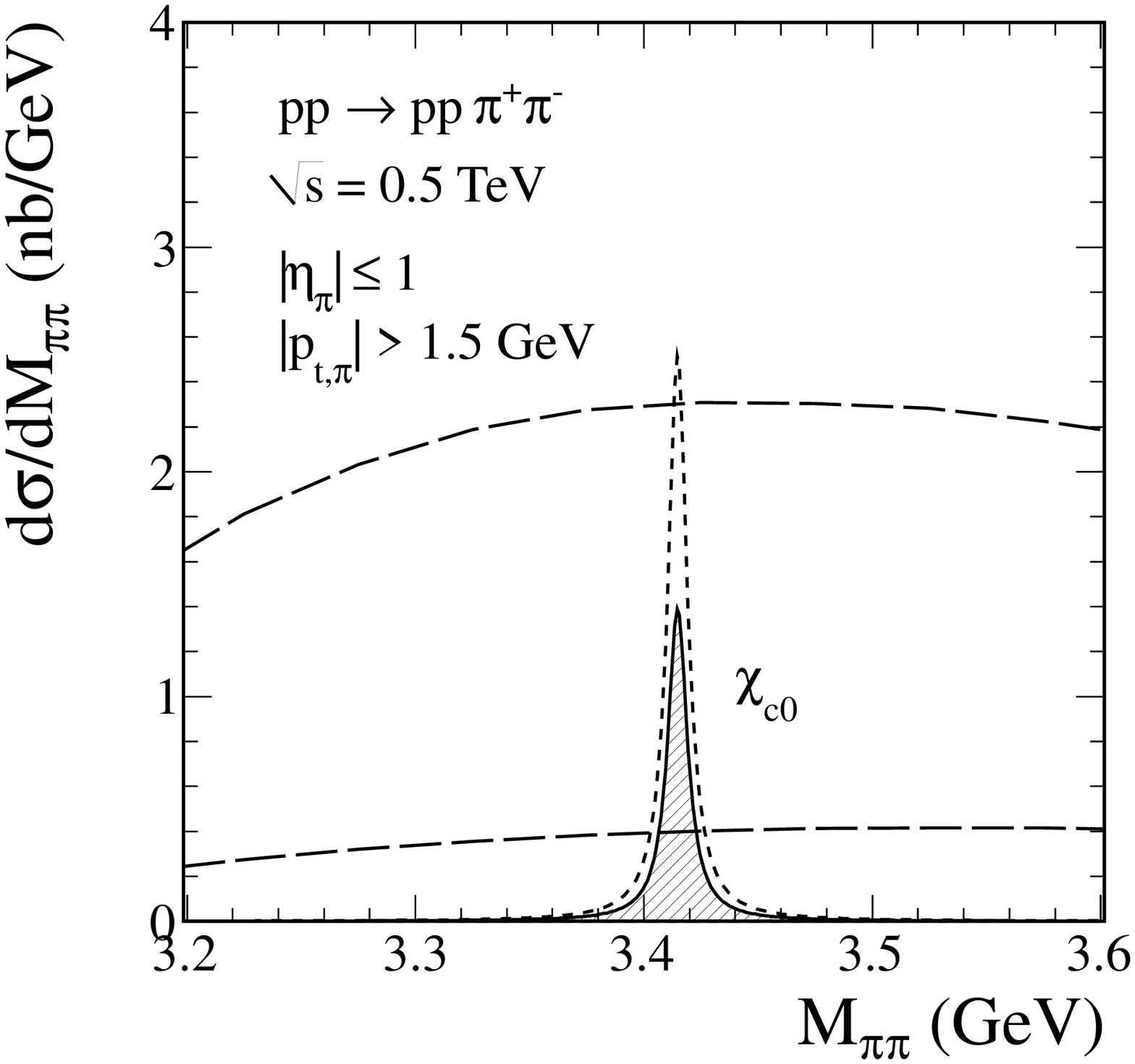}
\includegraphics[width = 0.32\textwidth]{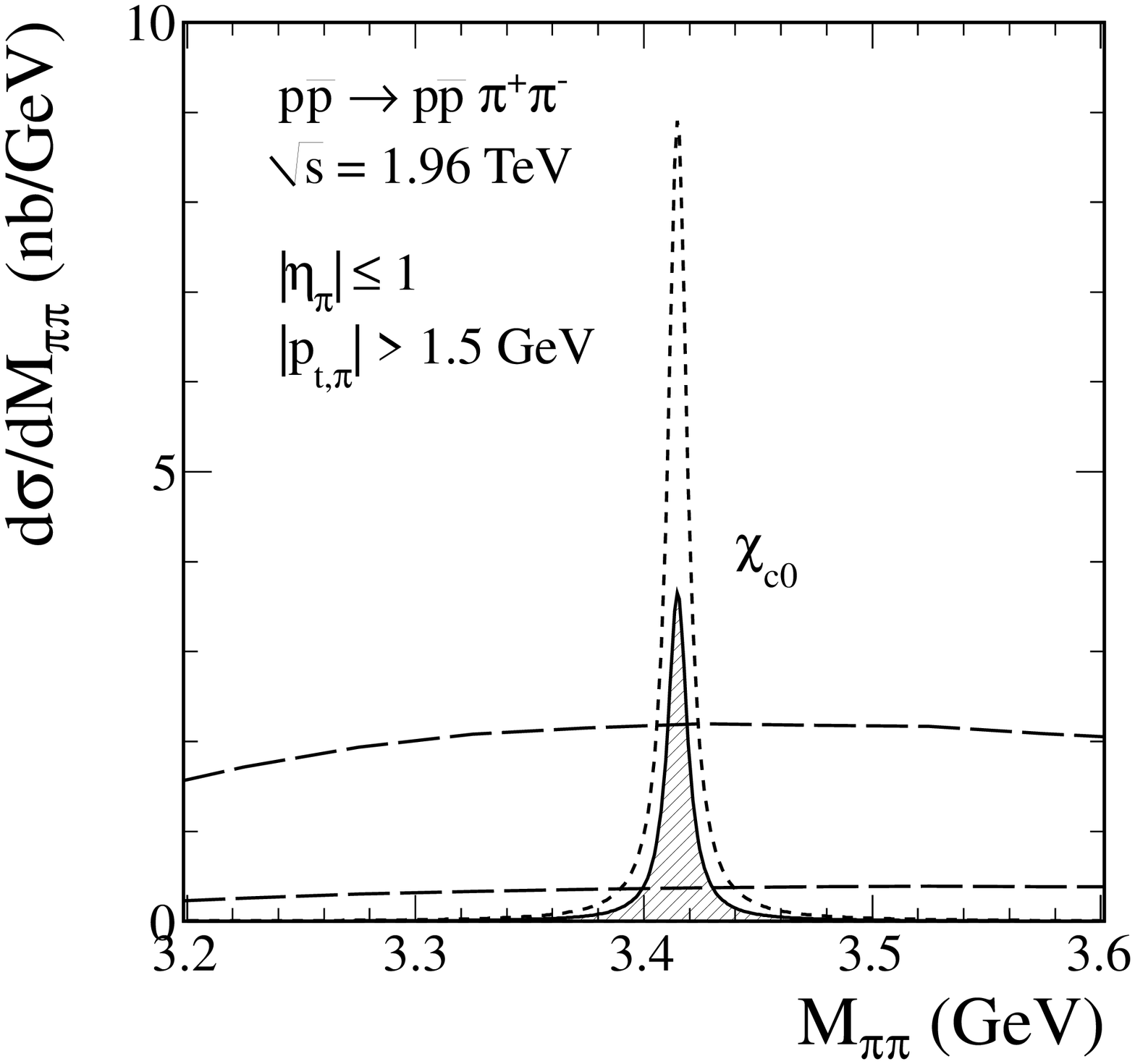}
\includegraphics[width = 0.32\textwidth]{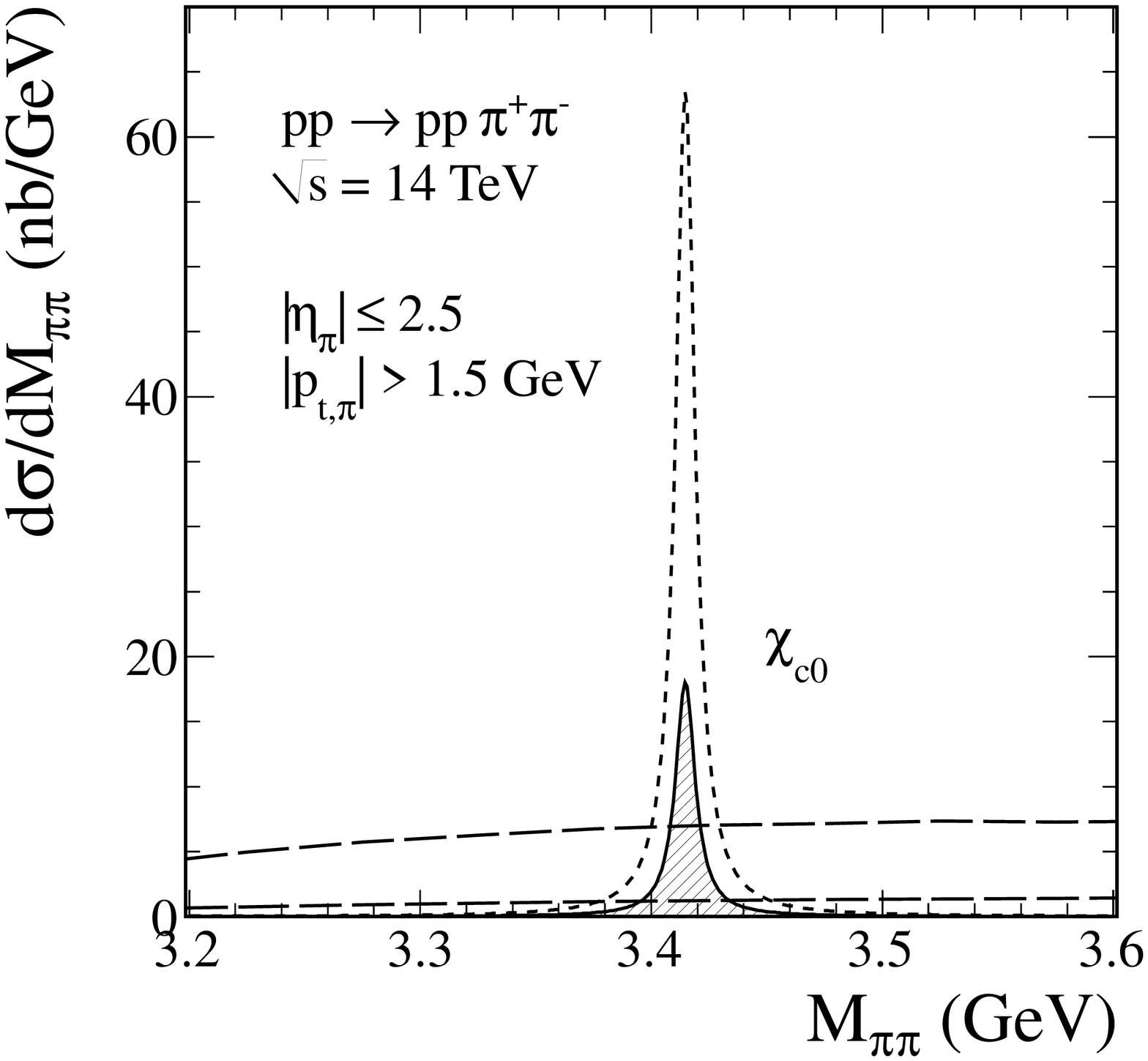}
  \caption{\label{fig:dsig_dmpipi_yptcut}
  \small
The $\pi^{+}\pi^{-}$ invariant mass distribution at $\sqrt{s} = 0.5,
1.96, 14$ TeV with the relevant restrictions in the pion
pseudorapidities and pion transverse momenta. The dashed lines
present the $\pi\pi$ continuum with the cut-off parameters
$\Lambda_{off, E}^{2}$ = 1.6, 2.0 GeV$^{2}$ (lower and upper dashed
lines, respectively). In calculating the $\chi_{c0}$ contribution we
use GRV94 NLO (dotted line) and GJR08 NLO (solid line) collinear gluon
distributions. The absorption effects for the $\chi_{c0}$ meson and
for the background were included. Clear $\chi_{c0}$ signal with
relatively small background for the Tevatron and LHC energies
can be observed when imposing extra cuts on $p_{t,\pi}$.}
\end{figure}

The question now is whether the situation can be improved by
imposing extra cuts. In Fig.~\ref{fig:dsig_dmpipi_yptcut} we show
results with additional cuts on both pion transverse momenta
$|p_{t,\pi}| > 1.5$ GeV. Now the signal-to-background ratio is
somewhat improved especially at the Tevatron and LHC energies. 
Shown are only purely theoretical predictions. In reality the situation is,
however, somewhat worse as both protons and, in particular, pion
pairs are measured with a certain precision which leads to an extra
smearing in $M_{\pi\pi}$. While the smearing is negligible for the
background, it leads to a modification of the Breit-Wigner peak for
the $\chi_{c0}$ meson
\footnote{An additional experimental resolution not included here can be taken
into account by an extra convolution of the Breit-Wigner shape with
an additional Gaussian function.}.
The results with more modern GJR UGDF are smaller
by about factor of 2-3 than those for somewhat older GRV UGDF.
%--------------------------------------------------------
%--------------------------------------------------------

In Table~\ref{tab:sig_tot_chic0} we have collected 
the numerical values of the cross sections 
(see $\sigma_{pp \to pp \chi_{c0}}$ in Eq.~(\ref{BW_form}))
for exclusive $\chi_{c0}$ production 
for some selected UGDFs at different energies.
%=============================================================================================
\begin{table}[h]
\caption{Integrated cross sections in nb (with absorption corrections)
for exclusive $\chi_{c0}$ production at different energies
with GRV94 NLO \cite{GRV} and GJR08 NLO \cite{GJR} collinear gluon distributions.
In this calculations we have taken
the relevant limitations in the pion pseudorapidities
$|\eta_{\pi}| < 1$ at the RHIC and Tevatron,
$|\eta_{\pi}| < 2.5$ at the LHC
and lower cut on both pion transverse momenta $|p_{t,\pi}|>$ 1.5 GeV.
}
\label{tab:sig_tot_chic0}
\begin{center}
\begin{tabular}{|c||c|c||c|c||c|c|}
\hline\hline
$\sqrt{s}$&\multicolumn{2}{c||}{full phase space}
          &\multicolumn{2}{c||}{with cuts on $\eta_{\pi}$}
          &\multicolumn{2}{c|}{with cuts on $\eta_{\pi}$ and $p_{t,\pi}$}\\
\cline{2-7}
(TeV)     &GRV   &GJR   &GRV  &GJR  &GRV  &GJR \\
\hline
0.5  & 63.6   & 34.5   & 14.2  & 7.9   & 7.3   & 4.0  \\
1.96 & 309.0  & 127.1  & 51.1  & 21.1  & 25.8  & 10.6 \\
14   & 1152.3 & 329.7  & 554.1 & 159.1 & 183.7 & 52.2 \\
\hline\hline
\end{tabular}
\end{center}
\end{table}
%=======================================================================================

\begin{figure}[!h]
\includegraphics[width = 0.32\textwidth]{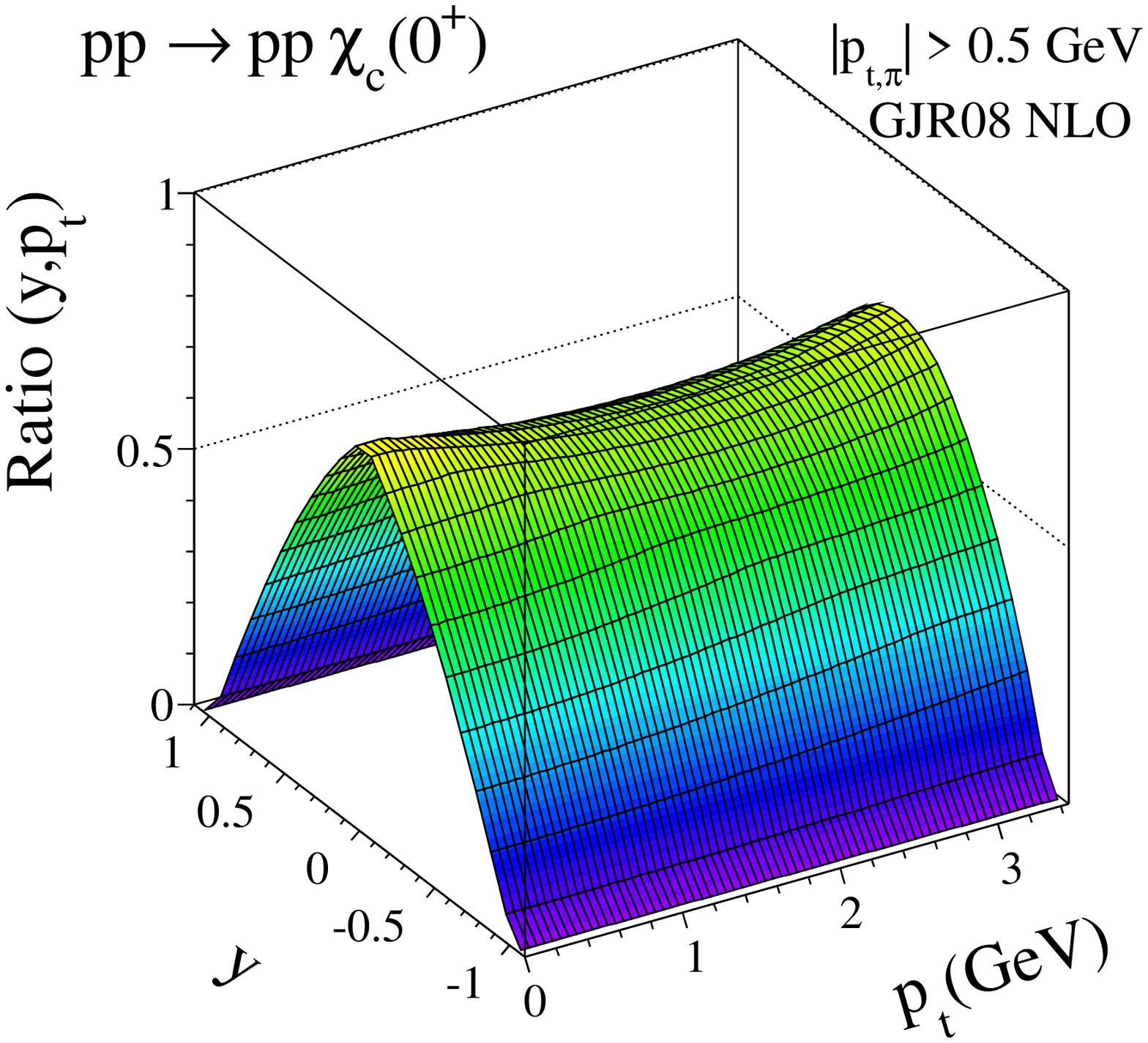}
\includegraphics[width = 0.32\textwidth]{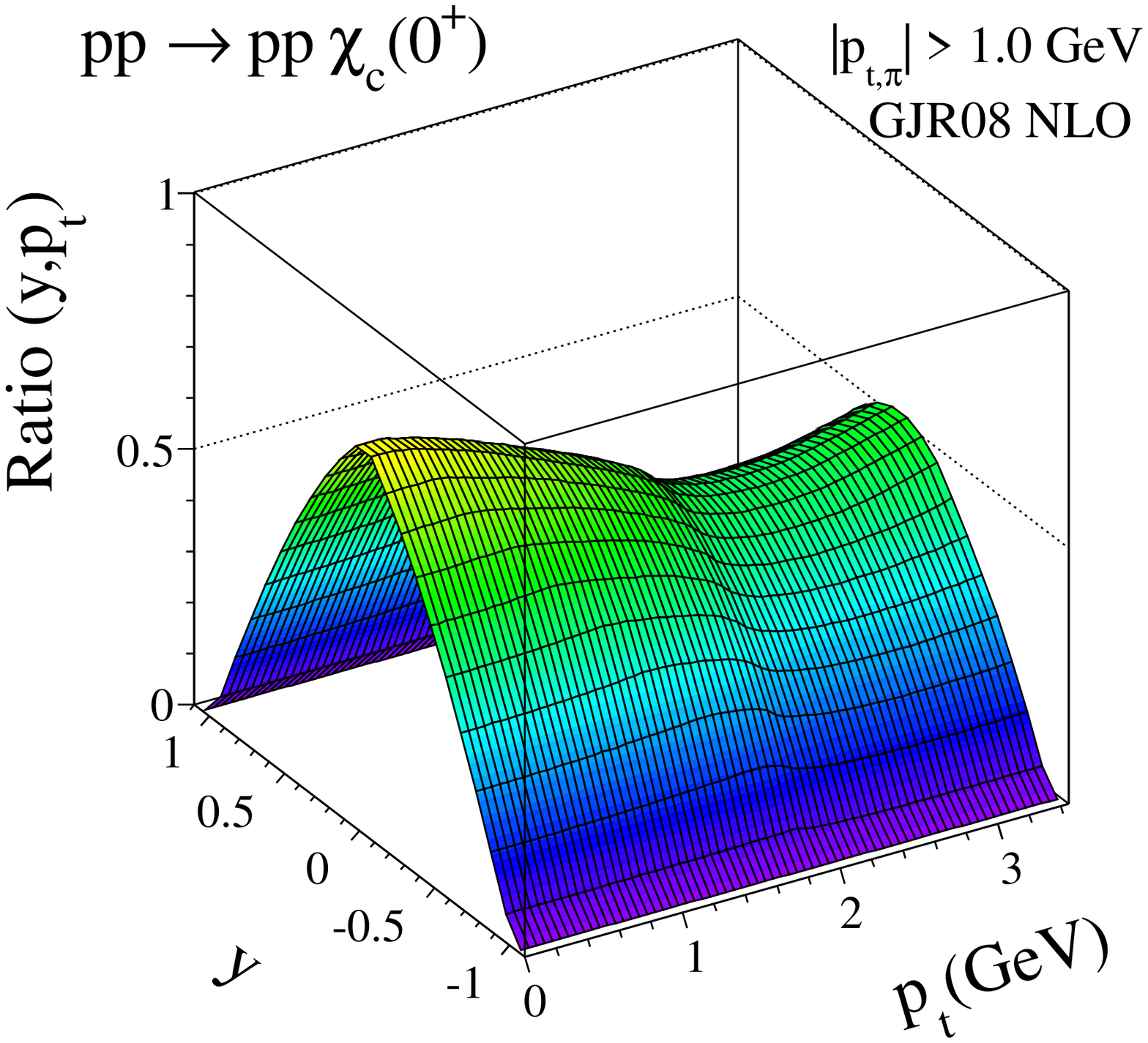}
\includegraphics[width = 0.32\textwidth]{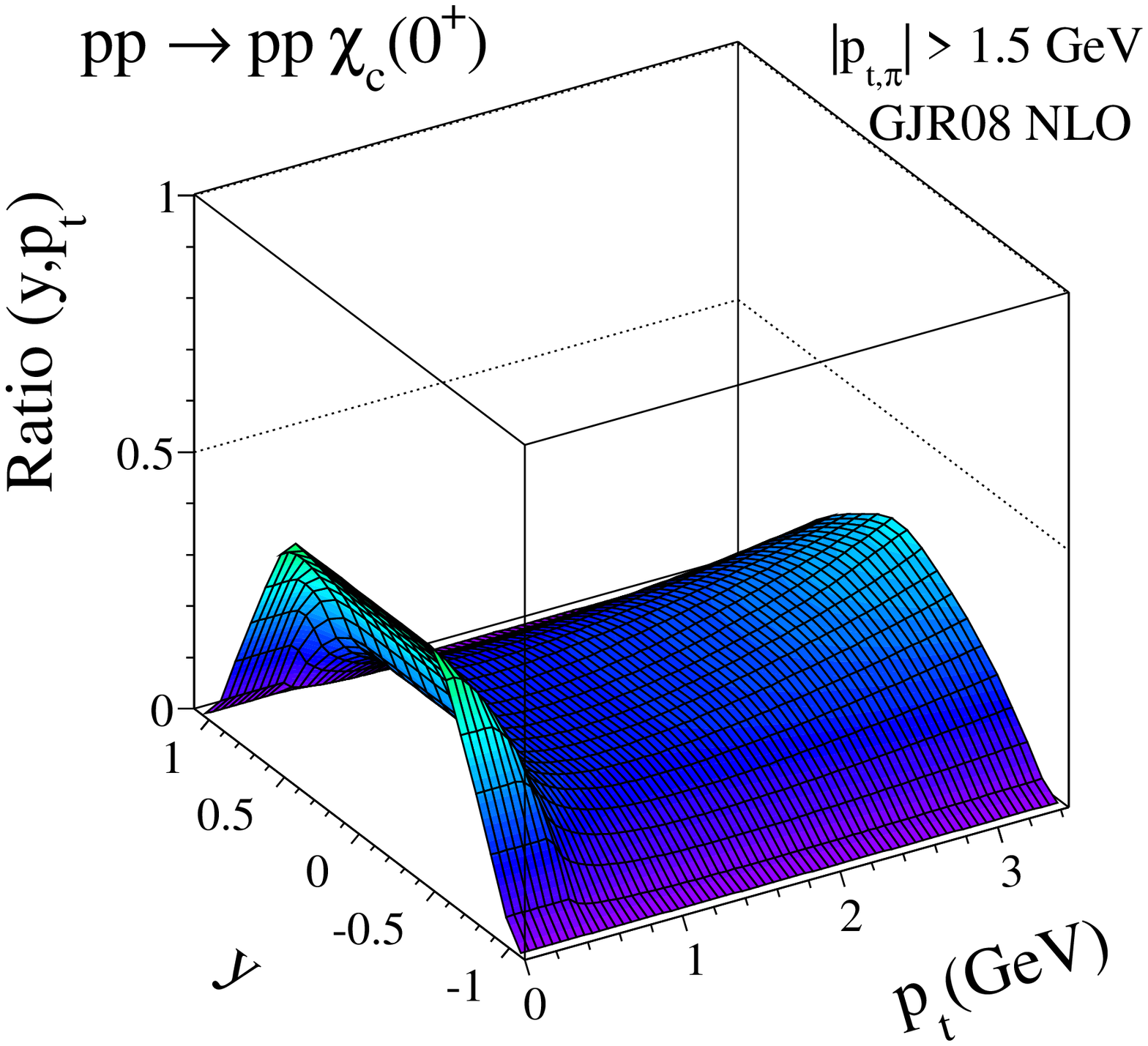}
\includegraphics[width = 0.32\textwidth]{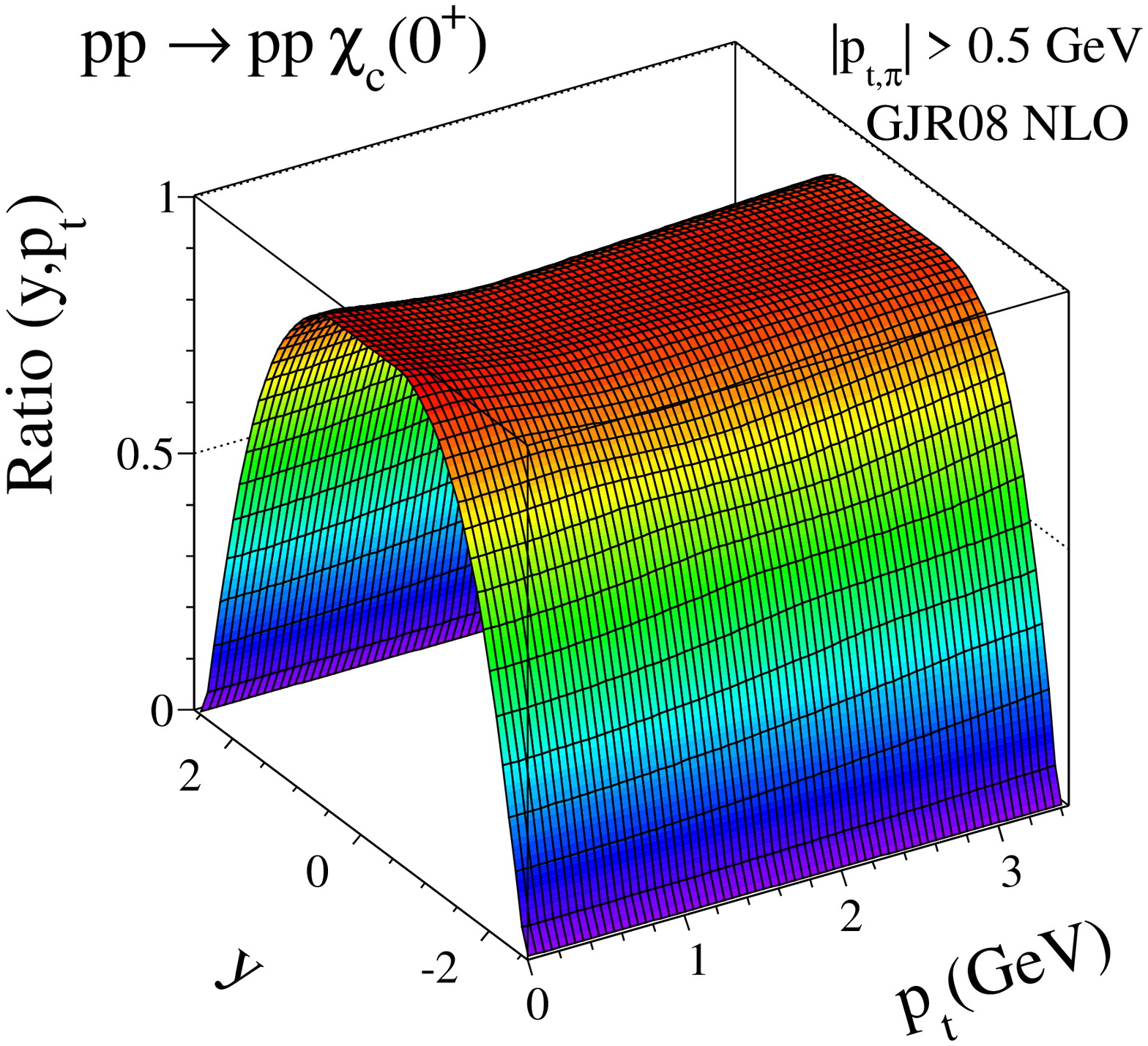}
\includegraphics[width = 0.32\textwidth]{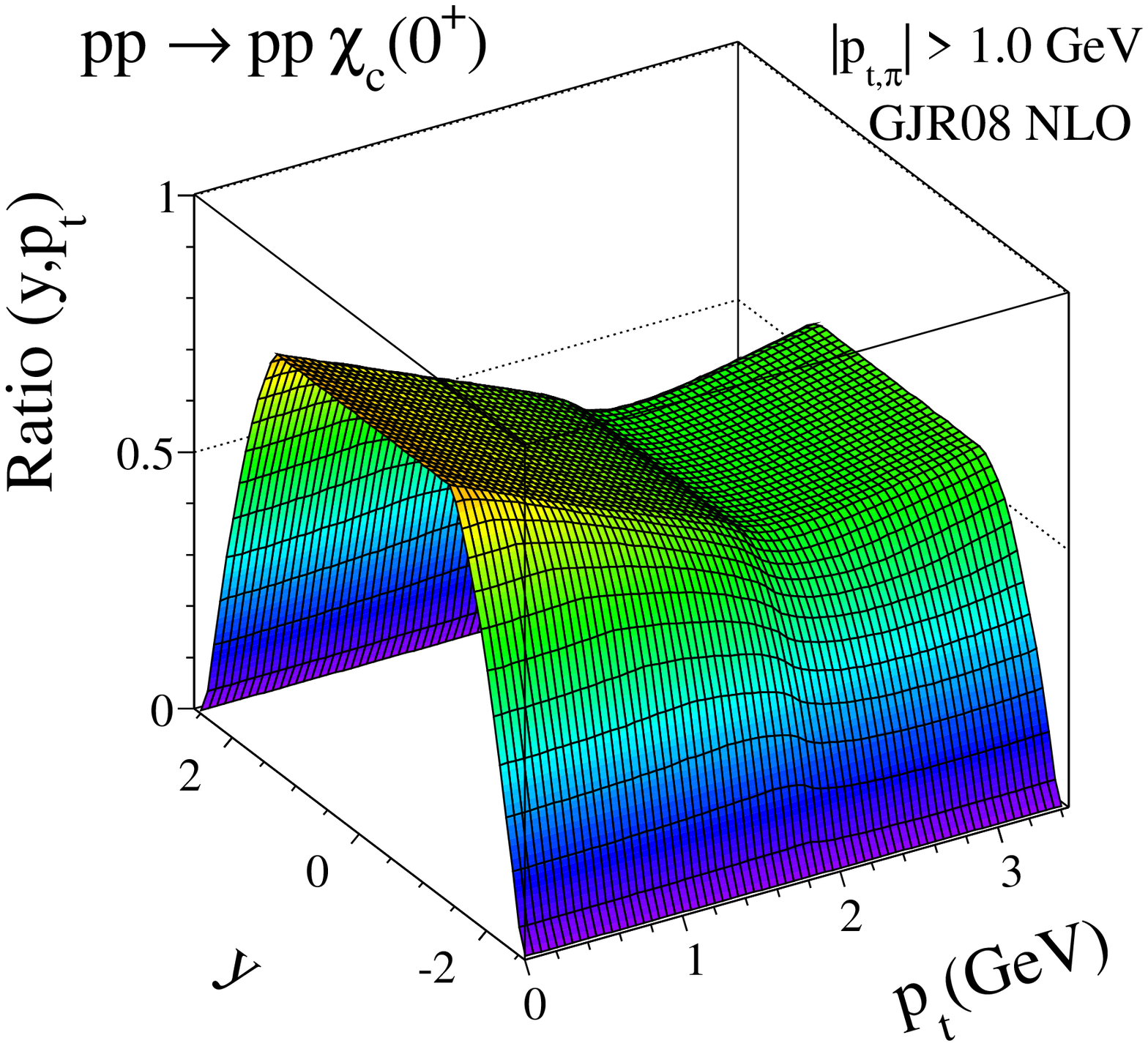}
\includegraphics[width = 0.32\textwidth]{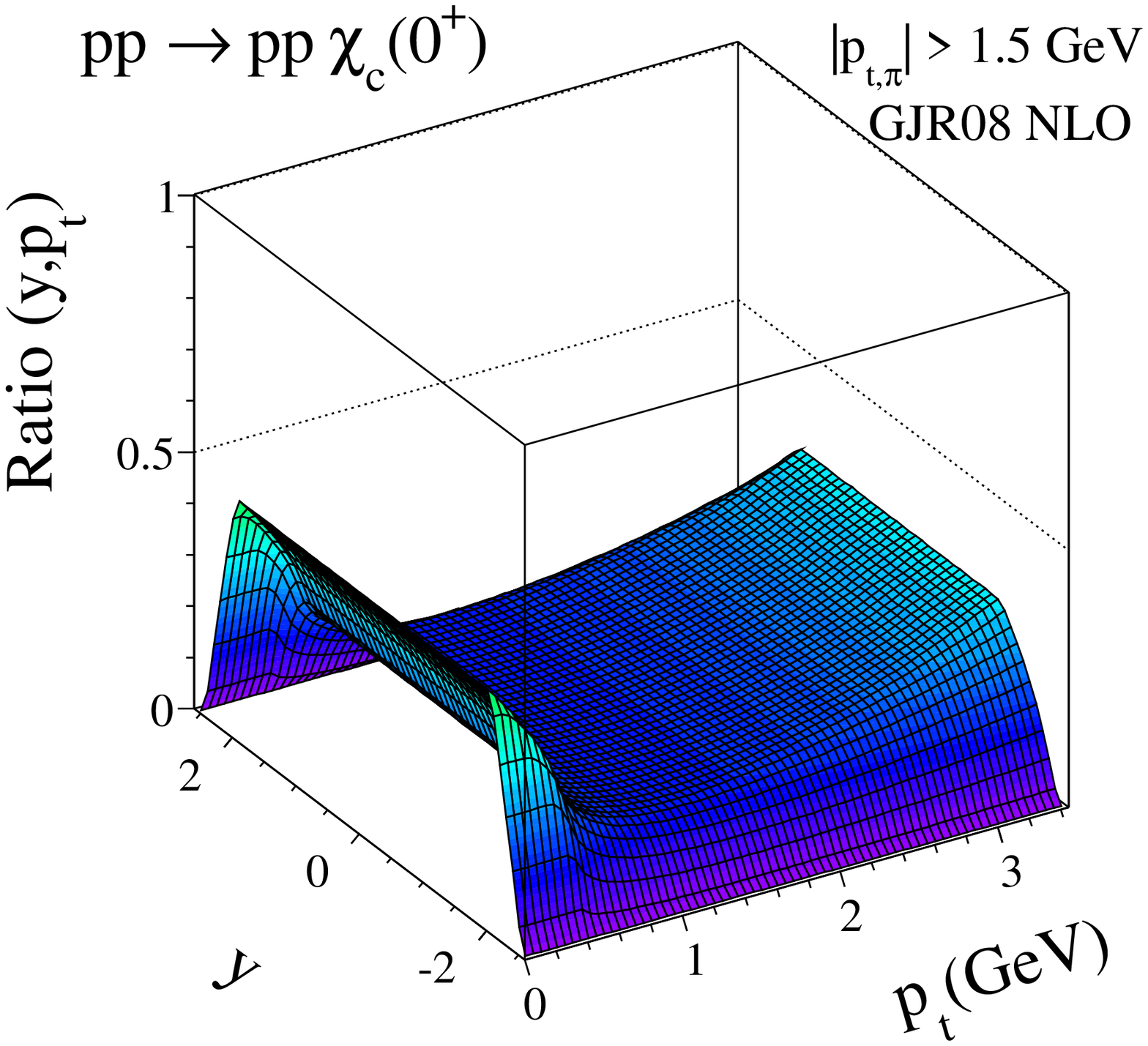}
  \caption{\label{fig:ratio}
  \small
Ratio of the two-dimensional cross sections in $(y,p_{t})$ for the
$pp \to pp \chi_{c0}$ reaction
%at $\sqrt{s} = 0.5, 1.96, 14$ TeV
with the relevant limitations on the pion pseudorapidities and a few
lower cuts on the pion transverse momenta $p_{t,\pi}$.
These calculations were done with GJR08 NLO \cite{GJR} UGDFs. 
The upper row is for the STAR detector ($|\eta_{\pi}| < 1$)
and the lower row for the ATLAS or CMS detectors ($|\eta_{\pi}| < 2.5$).}
\end{figure}

The main experimental task is to measure the distributions in the
$\chi_{c0}$ rapidity and transverse momentum. Can one recover such
distributions based on the measured ones
in spite of the severe cuts on pion kinematical
variables? In Fig.~\ref{fig:ratio} we show the two-dimensional ratio
of the cross sections for the $\chi_{c0}$ meson in its rapidity and
transverse momentum:
\begin{eqnarray}
\mathrm{Ratio}(y,p_{t}) =
\frac{d\sigma^{pp \to pp \chi_{c0}(\to \pi^{+} \pi^{-})}_{\mathrm{with \; cuts}} / dydp_{t}}
{d\sigma^{pp \to pp \chi_{c0}} / dydp_{t}}\,.
\label{ratio_formula}
\end{eqnarray}
The numerator includes limitations on $\eta_{\pi}$ and $p_{t,\pi}$.
These distributions provide a fairly precise evaluation of the expected
acceptances when experimental cuts are imposed.
The experimental data could be corrected by our two-dimensional
acceptance function to recover the distributions of interest.
%--------------------------------------------------------
%--------------------------------------------------------

%--------------------
\section{Conclusions}
\label{section:IV}
%--------------------

It was realized recently that the measurement
of exclusive production of $\chi_{c}$ via decay in the
$J/\psi + \gamma$ channel cannot give production cross sections for
different species of $\chi_{c}$.
In this decay channel the contributions of $\chi_{c}$ mesons
with different spins are similar and experimental resolution is not
sufficient to distinguish them.

In the present paper we have analyzed a possibility to measure the
exclusive production of $\chi_{c0}$ meson in the proton-(anti)proton
collisions at the LHC, Tevatron and RHIC via $\chi_{c0} \to
\pi^{+}\pi^{-}$ decay channel. 
Since the cross section for
exclusive $\chi_{c0}$ production is much larger than that for
$\chi_{c1}$ and $\chi_{c2}$ and the branching fraction to the $\pi \pi$ channel
for $\chi_{c0}$ is larger than that for $\chi_{c2}$ ($\chi_{c1}$
does not decay into two pions) the two-pion channel should provide
an useful information about the $\chi_{c0}$ exclusive production.

We have performed detailed studies of several differential
distributions and demonstrated how to impose extra cuts in order to
improve the signal-to-background ratio. The two-pion background was
calculated in a simple model with parameters adjusted to low energy
data. We have shown that relevant measurements at Tevatron and LHC
are possible. At RHIC the signal-to-background ratio is much worse
but measurements should be possible as well. Imposing cuts distorts
the original distributions for $\chi_{c0}$ in rapidity and
transverse momentum. We have demonstrated how to recover the
original distributions and presented the correction functions for
some typical experimental situations.

In the present paper we have concentrated on $\pi^{+} \pi^{-}$ channel only. 
This is mainly due to the relatively good control of the background. 
Measurements of other decay channels, e.g. $K^{+} K^{-}$, are possible as well 
and will be discussed elsewhere.
%We recall that hadronic channels
%are ideally suited for spin-parity analysis of the $\chi_{c}$ states.

\vspace{0.5cm}
%--------------------
{\bf Acknowledgments}
%--------------------

This study was partially supported by the Polish grant
of MNiSW N N202 249235.

%---------------------------------------------------------------------

\end{document}